\documentclass[conference]{IEEEtran}
\IEEEoverridecommandlockouts

\usepackage{graphicx}
\usepackage{float}
\usepackage{subcaption}
\usepackage{siunitx}
\usepackage{url}
\usepackage{cite}
\usepackage{multirow}
\usepackage{authblk}
\usepackage{hyperref}
\usepackage{etoolbox}

\newcounter{todonumber}
\stepcounter{todonumber}

\newcounter{writetodonumber}
\stepcounter{writetodonumber}

\newcolumntype{L}[1]{>{\raggedright\let\newline\\\arraybackslash\hspace{0pt}}m{#1}}
\newcolumntype{C}[1]{>{\centering\let\newline\\\arraybackslash\hspace{0pt}}m{#1}}
\newcolumntype{R}[1]{>{\raggedleft\let\newline\\\arraybackslash\hspace{0pt}}m{#1}}

\def\BibTeX{{\rm B\kern-.05em{\sc i\kern-.025em b}\kern-.08em
    T\kern-.1667em\lower.7ex\hbox{E}\kern-.125emX}}

\IEEEoverridecommandlockouts

\IEEEpubid{\makebox[\columnwidth]{979-8-3503-9973-8/23/\$31.00 ©2023 IEEE \hfill} \hspace{\columnsep}\makebox[\columnwidth]{ }}

\begin{document}
\title{Understanding the Benefits of Hardware-Accelerated Communication in Model-Serving Applications}

\author[a]{Walid A. Hanafy}
\author[b]{Limin Wang}
\author[b]{Hyunseok Chang}
\author[b]{Sarit Mukherjee}
\author[b]{T. V. Lakshman}
\author[a]{Prashant Shenoy}
\affil[a]{University of Massachusetts Amherst}
\affil[b]{Nokia Bell Labs}

\maketitle
\IEEEpubidadjcol

\begin{abstract}
It is commonly assumed that the end-to-end networking performance of edge offloading is purely dictated by that of the network connectivity between end devices and edge computing facilities, where ongoing innovation in 5G/6G networking can help. However, with the growing complexity of edge-offloaded computation and dynamic load balancing requirements, 
an offloaded task often goes through a multi-stage pipeline that spans across multiple compute nodes and proxies interconnected via a dedicated network fabric within a given edge computing facility. As the latest hardware-accelerated transport technologies such as RDMA and GPUDirect RDMA are adopted to build such network fabric, there is a need for good understanding of the full potential of these technologies in the context of computation offload and the effect of different factors such as GPU scheduling and characteristics of computation on the net performance gain achievable by these technologies.
This paper unveils detailed insights into the latency overhead in typical machine learning (ML)-based computation pipelines and analyzes the potential benefits of adopting hardware-accelerated communication. To this end, we build a model-serving framework that supports various communication mechanisms. Using the framework, we identify performance bottlenecks in state-of-the-art model-serving pipelines and show how hardware-accelerated communication can alleviate them. For example, we show that GPUDirect RDMA can save 15--50\% of model-serving latency, which amounts to 70--160 ms.
\end{abstract}


\section{Introduction}\label{sec:intro}

The concept of edge computing was pioneered more than a decade ago~\cite{satya09,bonomi}, and yet the role of edge computing remains critical even today because functional requirements and user expectations for applications have constantly been surpassing even the most sophisticated on-device capabilities~\cite{mahadev21}.  For example, multi-user cloud gaming, machine learning (ML)-based wearable cognitive assistance, immersive 360-degree point cloud video streaming, industrial robot control, etc. heavily rely on geographically close-by compute resources, accessible to end devices via low-latency, high-throughput interconnects. In making edge computing a reality, there have been two important industry trends. On the networking side, the advances in 5G/6G technologies have been instrumental in enabling offloaded computation and associated data delivery to meet stringent latency/throughput requirements.  On the computation side, the arrival of new chip technologies (e.g., GPU, TPU) has been a catalyst for accelerating and scaling required computation within server hardware.

As more and more practical use cases of edge offloading are emerging, driven by these trends, the research community has been dedicating significant research efforts to optimize the latency performance of edge offloading within a given edge computing infrastructure. There have been works on utilizing adaptive computation for low latency~\cite{clipper,deepslicing,naveen21}, and proposing intelligent workload scheduling in compute clusters or a single node~\cite{qliang2022,clockwork,msml}. 
Although the existing works differ in their approaches and scopes, one commonality they share is that the primary focus is on compute resources or computation itself, but not on the underlying networking, in particular, \emph{the network fabric within an edge computing infrastructure}.  A common assumption is that the end-to-end networking performance of edge offloading is purely dictated by that of the network connectivity between end devices and edge computing facilities, where ongoing innovation in 5G/6G networking can help.  However, with the growing complexity of offloaded computation and dynamic load balancing requirements within a single edge domain, an offloaded task often goes through a multi-stage pipeline which spans across multiple compute nodes and proxies interconnected via a dedicated network fabric within a given edge computing infrastructure.  Therefore, the performance of such internal network fabric and its interaction with task execution can also play a nontrivial role in the end-to-end latency of edge offloading.

The latest hardware-accelerated transport technologies such as Remote Direct Memory Access (RDMA) and GPUDirect are suitable for building the network fabric for interconnecting compute nodes in current edge computing environments~\cite{edgeservice}.
Unlike TCP/IP-based communication, where a server CPU is involved in packetizing and transferring data via the operating system's protocol stack, RDMA and GPUDirect bypass the server CPU and the operating system, and directly write data into a destination processor's memory (it could be CPU memory or GPU memory).  This remote zero-copy mechanism allows these technologies to decrease data transfer latency and increase service throughput, presenting a ``lower-bound'' latency to other remote computation offloading techniques.
However, there is still a lack of understanding on the full potential of the existing hardware-based transport technologies in the context of computation offload and the effect of different factors (e.g., GPU scheduling, computation type and size) on the net performance gain achievable by them.

This paper aims to unveil detailed insights into the performance overhead of typical model-serving pipelines, which are often hard to get from existing feature-rich model-serving systems, and in turn, to highlight potential performance gains that could be achieved by adopting hardware-accelerated transport in different deployment scenarios.
To achieve this goal,
we build a model-serving application framework 
with support for different communication mechanisms (e.g., TCP, RDMA) and with the capability to provide fine-grained visibility into model-serving pipeline stages.\footnote{The source code for the model-serving system is available at \url{https://github.com/nokia/accelerated-offloading}.}
Such exploratory features are not available in existing off-the-shelf model-serving systems. We use the framework to explore a wide-range of scenarios that resemble real-world edge deployments.

Our systematic evaluation demonstrates that hardware-accelerated transport, in particular GPUDirect RDMA (GDR), presents a promising approach to build low-latency edge offloading infrastructures, saving 15--50\% of model-serving latency, which translates to \SIrange[range-phrase=--,range-units=single]{70}{160}{\ms}, compared to TCP-based transport in a wide range of setups. This study helps us understand the benefits of hardware-accelerated communication in model-serving applications, as summarized below.
\noindent\textbf{(1) Communication fraction matters.} Hardware-accelerated transport provides the most benefit when communication takes a significant fraction of time in a given model-serving pipeline. 
With increasing GPU processing capabilities and application network I/O requirements, the proportion of communication overhead is expected to become more significant.

\noindent\textbf{(2) Protocol translation is worthwhile.} Adopting hardware-accelerated transport within a given compute cluster can provide substantial latency benefit compared to end-to-end TCP pipelines, even at the cost of protocol translation.

\noindent\textbf{(3) Data copies are bottlenecks.} Host-to-device (H2D) and device-to-host (D2H) copies can quickly become a bottleneck as concurrency increases within a GPU. Issuing copy commands interferes with execution in a GPU. GDR can alleviate these problems by skipping the GPU copy queues all along.

\noindent\textbf{(4) Effectiveness of prioritization is limited due to copy-engine's coarse granular interleaving.} GPU copy-engine's coarse granular interleaving limits the ability of high-priority clients to prioritize their execution over other clients.

\section{Background}\label{sec:background}
In this section, we provide an overview of the technologies we evaluate in this paper.

\subsection{Edge Offloading}

\begin{figure}[t]
    \centering
    \vspace{-2ex}
    \includegraphics[width=0.45\textwidth]{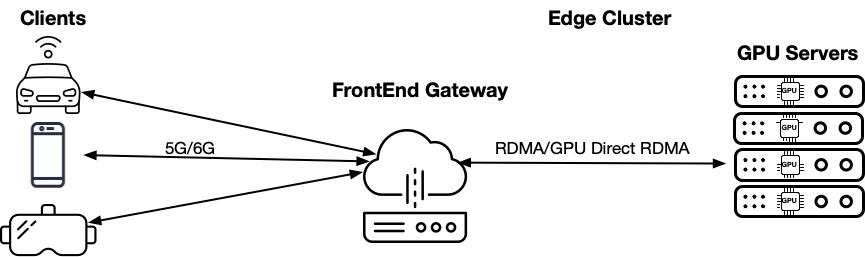}
    \caption{Edge offloading architecture.}
    \label{fig:edge}
    \vspace{-3.5ex}
\end{figure}

In a typical edge offloading architecture (Figure~\ref{fig:edge}), end devices offload computational tasks (e.g., object recognition in a camera view, collision-free robot navigation) to a nearby edge computing facility via request/response transactions.
When an end device requests for an offloading service, it submits corresponding data to a frontend gateway of an edge computing facility over existing access networks. The gateway then dispatches the request to compute servers available at the facility over an internal network fabric. Once the required computation gets executed on the data, a response is generated and sent back to the requesting device via the gateway. Unlike the external access networks interconnecting end devices and edge computing facilities, the edge-internal network fabric is
under the control of a given edge computing facility, and can be enhanced by leveraging the latest hardware-accelerated network fabric technologies such as RDMA and GPUDirect. 
In the following, we describe how an edge offloading task can be accelerated within a given edge computing facility via RDMA and GPUDirect RDMA.
\subsection{Edge Offloading over RDMA}
RDMA is a hardware mechanism through which a local peer can directly access a remote peer’s memory without the intervention of the remote peer's CPUs and the network stack traversal overhead.  RDMA was originally designed to interconnect high-performance computing (HPC) clusters on specialized high-throughput, low-latency InfiniBand (IB) networks. Many of today's RDMA deployments are based on RoCEv2 (RDMA over Converged Ethernet), where packets are encapsulated in UDP/IP packets and carried over the commodity Ethernet fabric. Figure \ref{fig:zero_copy_steps}(a) depicts a typical workflow of a compute-intensive request utilizing a GPU on a remote server over RDMA. When the request along with necessary data arrives at the server's RDMA-capable NIC (RNIC) as RDMA traffic, it gets DMAed to the server's RAM (steps 1--2). Then it gets copied from RAM to GPU memory and serviced in the GPU, after which the result gets copied back to RAM (steps 3--5). As a response, the result is DMAed from RAM to the RNIC and sent out as outgoing RDMA traffic by the RNIC (steps 6--7). Throughout the workflow, the CPUs of the server may issue control plane instructions at certain steps (e.g., steps 3--6), but is not directly involved in data movement.

\subsection{Edge Offloading over GPUDirect RDMA}

\begin{figure}[t]
\centering
    \begin{subfigure}{0.45\linewidth}
        \centering
       \includegraphics[width=\textwidth]{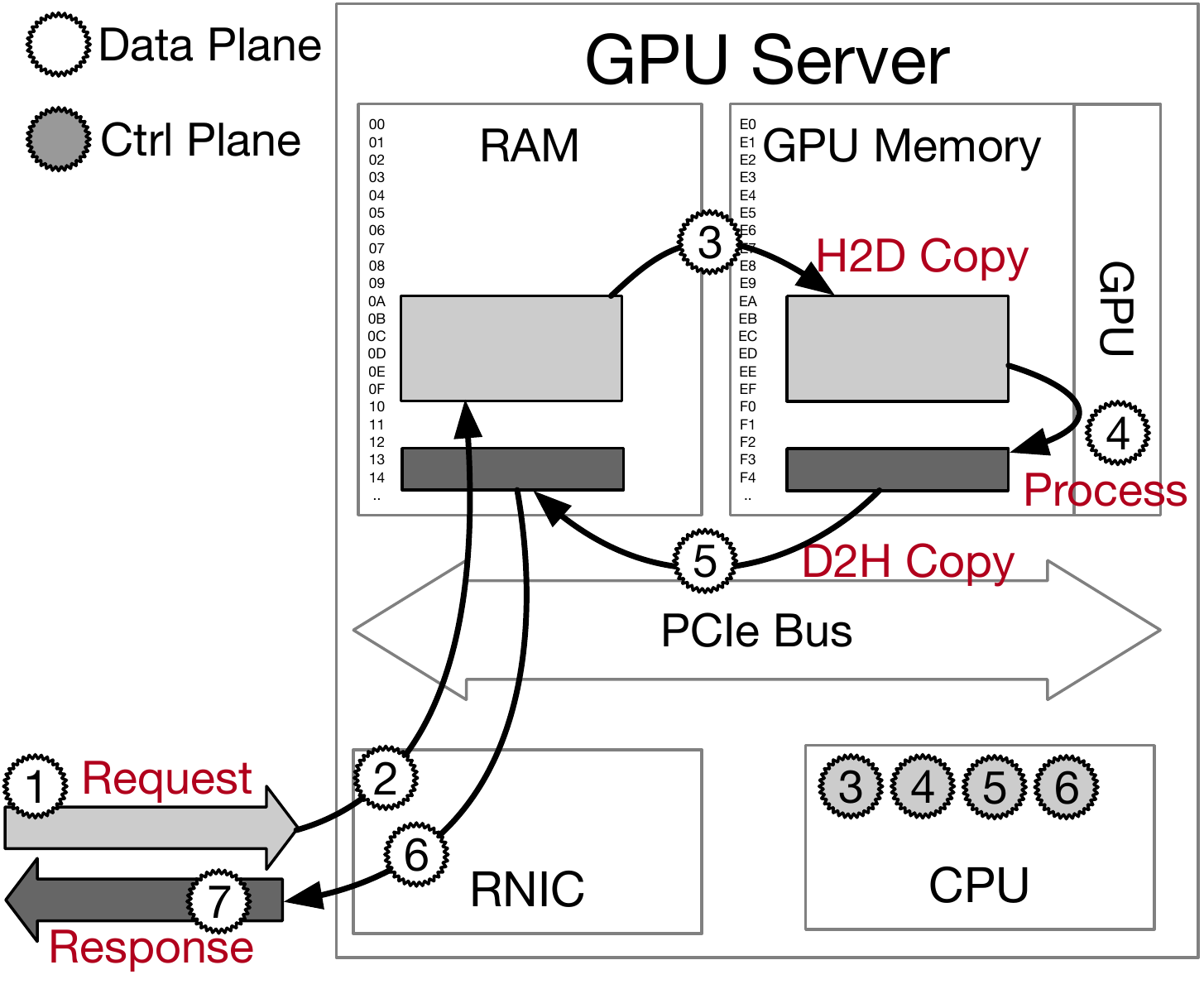}
       \caption{RDMA}
       \label{fig:rdma_steps}
    \end{subfigure}
\hfill
    \begin{subfigure}{0.48\linewidth}
        \centering
        \includegraphics[width=\textwidth]{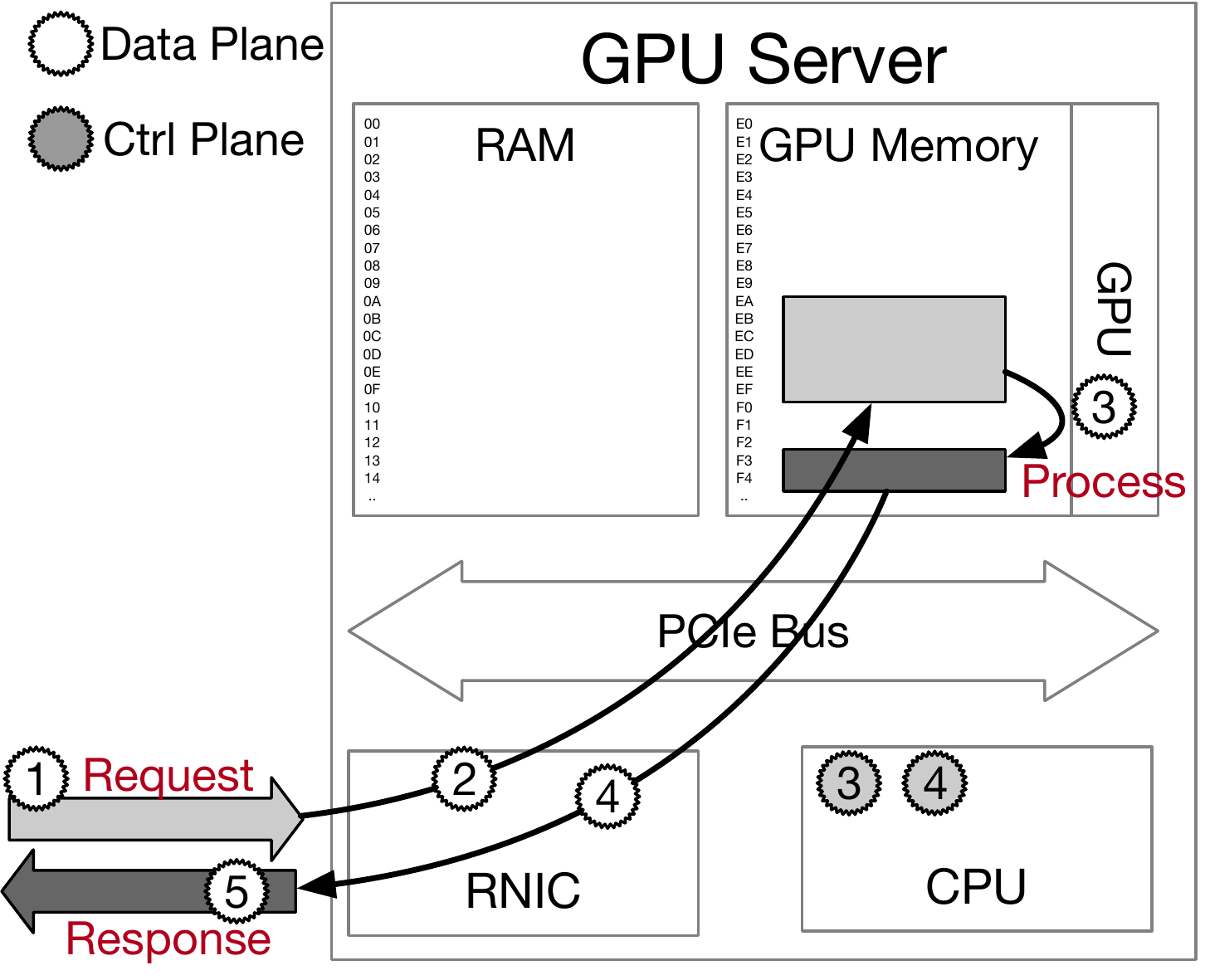}
        \caption{GDR}
        \label{fig:gpudirect_steps}
    \end{subfigure}
\vspace{-1.5ex}
\caption{Request-response transaction over RDMA/GDR.}%
\label{fig:zero_copy_steps}
\vspace{-3ex}
\end{figure}

GPUDirect is a suite of technologies introduced by NVIDIA to enhance data movement and access for their GPUs.
In particular, GPUDirect RDMA (GDR) enables PCIe devices like RNICs to directly access GPU device memory. This eliminates the involvement of CPUs and the staging buffer copies of data via main memory for inter-node GPU communication, thereby reducing CPU overhead and improving latency. To support GDR, NVIDIA provides an operating system extension that enables DMA bus mapping of GPU device memory to allow GPU memory to be directly used as RDMA target memory regions just like normal main memory.
With GDR, leveraging a remote GPU for an edge offloading task can be substantially simplified, as shown in Figure \ref{fig:zero_copy_steps}(b). Since GPU device memory can be directly employed as RDMA target memory regions, an incoming service request and its data can be DMAed by the RNIC to GPU memory without going through system RAM (steps 1--2), and GPU processing can take place right away (step 3). Similarly, the service result can also be directly DMAed from GPU memory to RNIC for output (steps 4--5). The copy in/out of data between RAM and GPU memory in Figure \ref{fig:zero_copy_steps}(a) is completely avoided. 

\subsection{GPU Scheduling}\label{sec:bg_gpu}

Typical GPU hardware contains an array of multi-threaded execution engines called Streaming Multiprocessors (SMs), as well as a multi-level memory system and dedicated copy engines that copy data to and from host memory over the PCIe bus, in parallel to execution engines. Each SM comprises a large number of processing cores known as CUDA cores in the case of NVIDIA GPUs.
NVIDIA GPUs are programmed on the CUDA platform, which presents a programming model and easy-to-use APIs for utilizing available CUDA cores. A typical CUDA program is composed of multiple kernels (functions) designed to exploit the parallel processing of CUDA cores.  Kernels are grouped into blocks, where each block contains multiple threads.

Although GPU scheduling algorithms on NVIDIA hardware platforms are proprietary, researchers have been trying to gain insights on them~\cite{tx2scheduling2017, yang2018avoiding}. In summary, to issue kernels or to allocate memory, programs create a CUDA context which communicates with the GPU driver and holds the execution state. CUDA contexts issue kernels to a default stream called \emph{NULL} stream. For higher GPU utilization and faster execution, it is possible to use multiple streams. 
With multiple streams, the execution engine schedules kernels' blocks across streams in a priority-accommodating round-robin fashion~\cite{tx2scheduling2017, yang2018avoiding}. Once kernels are in the execution engine queue, their blocks are scheduled in an FCFS fashion, where each block is launched based on the availability of CUDA cores and memory. CUDA streams also support priority-based scheduling, where different CUDA streams have different priorities. The priorities affect the execution on the block level in a non-preemptive way.  Another way of sharing GPUs is by using multiple contexts over multiple threads or processes. In this case, GPU execution engines are shared among contexts in a time-sliced fashion.
Finally, NVIDIA GPUs support Multi-Process Service (MPS), which allows packing threads from multiple contexts to be running at the same time~\cite{mps}. This resembles the execution of multiple streams without potential head-of-queue blocking in streams.
Although all these resource sharing methods increase system efficiency at scale, they come with the cost of lower performance predictability. Researchers have developed methods to cope with the unpredictability~\cite{qliang2022}, or improve predictability by disabling sharing methods \cite{clockwork}. However, the trade-off between predictability and utilization has not been fully explored.
\section{Methodology}\label{sec:method}

In this paper, we seek to investigate the role of hardware-accelerated network fabric in enabling low-latency computation offload at the edge.  Given that ML model-serving is a popular type of edge-offloaded computation, we focus on model-serving offload.  In order to study the complex interplay between network fabric technologies and the rest of model-serving pipelines and derive generalizable findings, what is needed is somewhat open-ended deployment environments, where we can easily enable different features and analyze their impact. Off-the-shelf model-serving systems~\cite{triton, tf-serving} are not suitable in this respect due to the following reasons. First of all, the existing systems only support TCP-based application protocols such as HTTP and GRPC, but do not support hardware-accelerated transport primitives such as RDMA and GDR. Second, as deployment-ready systems, they do not come with fine-grained profiling and tracing capability that will help us understand performance bottlenecks in model-serving pipelines.
Finally, production-grade systems often come with various add-on features such as security, clients priority, etc., which may not help with or even complicate our investigations. These factors motivate us to build our own model-serving system from scratch that is flexible and generic enough to serve as a reference model-serving testbed. In the following, we describe the functional details of our model-serving system.

\subsection{Model-Serving Pipeline Stages}

\begin{figure}[t]
    \centering
    \includegraphics[width=0.45\textwidth]{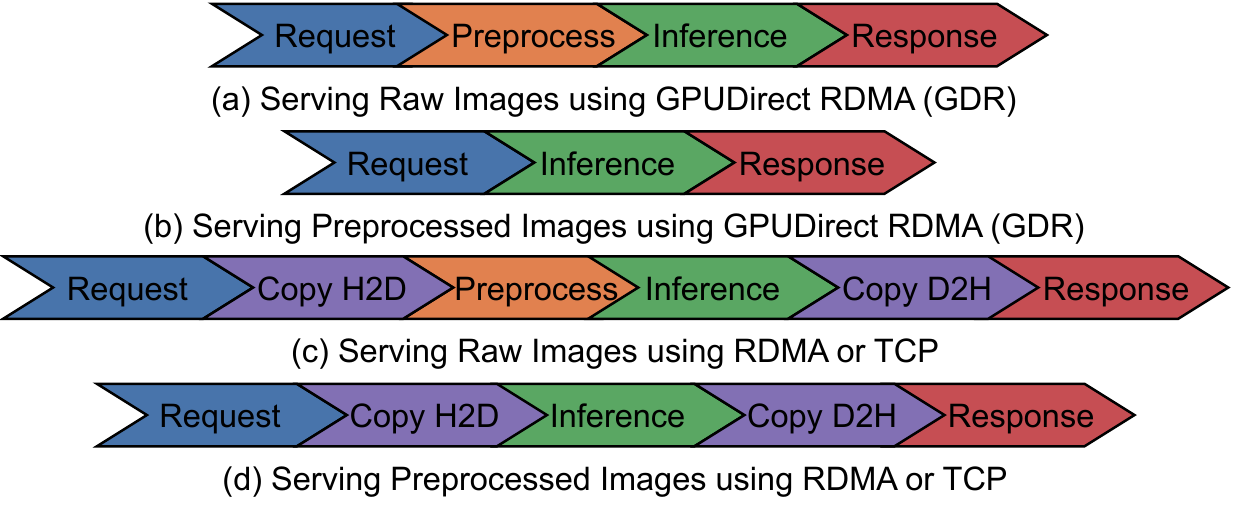}         
    \vspace{-1ex}
    \caption{Model-serving stages.}
    \label{fig:model_serving_steps}
\vspace{-4ex}
\end{figure}

In our framework, a model-serving pipeline consists of request handling, preprocessing, inference, and response handling stages. The preprocessing stage ensures that client-submitted data is compatible with the model requirements (e.g., input size) in case that the client submits raw data. The inference stage executes a given model with client data. The request/response handling stages proceed differently based on the underlying transport mechanism, as explained below.

To support model-serving over RDMA, a server and a client first go through an RDMA-specific connection setup procedure.  During this time, they create a set of queues for sending/receiving data over RDMA and for receiving work completion events, allocate memory buffers to hold request and response data, and exchange connection-related metadata~\cite{rdmadoc}.  Once a connection is created, the client sends a model-serving request to the server by posting a work request (WR) to its send queue, and blocks until it receives work completion (WC) events for the request as well as for a corresponding response from the server.  We use \texttt{RDMA\_WRITE} operation for both the request and the response.
When the server receives a WC event for the request, as shown in Figure~\ref{fig:zero_copy_steps}(a), it first copies the data from the client's request buffer to the GPU memory using \texttt{cudaMemcpy} with \texttt{cudaMemcpyHostToDevice} flag. Then it processes the request according to the application requirements. Lastly, it copies the data back to the client's response buffer, this time with \texttt{cudaMemcpyDeviceToHost} flag, and pushes a WR to its send queue and waits for a WC event. GDR follows the same steps as RDMA, except that, on the server, we allocate GPU memory rather than host memory, and that H2D and D2H copies are omitted (Figure~\ref{fig:zero_copy_steps}(b)). 

For TCP-based transport, we choose ZeroMQ~\cite{zeromq} over HTTP and GRPC for the following reason. The RDMA-based transport in our framework allows data to be transmitted with memory read/write semantic, and hence does not incur data (de)serialization overhead.  Whereas common TCP-based protocols like HTTP and GRPC require data (de)serialization, and therefore comparing end-to-end latency between HTTP/GRPC and RDMA is not fair.  Unlike HTTP/GRPC, ZeroMQ does not require data serialization, and hence it can be a fair comparison with RDMA protocol. We use a Router-Dealer proxy where the server allocates the same number of threads as the number of clients. Each thread reuses its memory buffers to avoid memory allocation overheads.
Figure \ref{fig:model_serving_steps} summarizes the model-serving pipelines for different communication mechanisms. The difference lies in the processing stages for raw and preprocessed data and the steps on the server side where data copies are selectively needed.

\subsection{Performance Metrics}

\begin{table}[t]
\centering
\caption{Performance metrics.}
\label{tab:metrics}
\resizebox{0.9\columnwidth}{!}{%
{\setlength\tabcolsep{0.5ex}
\begin{tabular}{|C{1.2cm}|C{3cm}|C{4cm}|}
    \hline
    \textbf{Category} & \textbf{Metric} & \textbf{Description}\\ \hline \hline
    & \texttt{total-time} & End-to-end model-serving latency \\\hline
    \multirow{2}{1cm}{Transport} &  \texttt{request-time} & Time taken to send a request \\\cline{2-3}
    & \texttt{response-time} & Time taken to send a response\\\hline
    & \texttt{copy-time} & H2D copy time + D2H copy time \\\cline{2-3}
    GPU & \texttt{preprocessing-time} & Time taken in preprocessing \\\cline{2-3}
    & \texttt{inference-time} & Time taken in model inference \\\hline
    CPU & \texttt{cpu-usage} & CPU usage in user and kernel\\\hline
    Memory & \texttt{memory-usage} & RAM and GPU memory usage\\
    \hline
\end{tabular}%
}
}
\vspace{-2ex}
\end{table}

To understand the performance bottlenecks of a model-serving pipeline, fine-grained visibility into the pipeline is required. Thus, we enable detailed time profiling for individual pipeline stages in the model-serving system.
The client-perceived end-to-end model-serving latency is broken into two components: (i) transport latency and (ii) GPU latency.  
Within GPU latency, \texttt{copy-time} is only applicable to TCP/RDMA-based communication as GDR moves data directly to GPU memory. 
To measure GPU-related latency components, we inject CUDA events between steps and measure the time between the events.
Transport delay components \texttt{request-time} and \texttt{response-time} capture client-to-server and server-to-client communication overhead, respectively, which are measured differently for different transport methods. 
For RDMA/GDR-based transport, which is offloaded to an RNIC, we measure the delay as the time between posting an RDMA WR and receiving a corresponding WC event. TCP-based ZeroMQ communication overhead is measured with processing time for \texttt{zmq\_send}() API on server response, while the request time is the time difference  between the \texttt{total-time} and total server time.
Besides measuring model-serving latency, we also capture CPU/memory resource usages using Linux \texttt{/proc} file system and \texttt{nvidia-smi}. All reported performance metrics are summarized in Table~\ref{tab:metrics}. The metrics are collected with a varying number of clients, where each client sends 1000 requests in a closed-loop fashion. 

\subsection{Experimental Scenarios}
 
The flexibility of our model-serving system allows us to evaluate model-serving pipelines across a wide range of deployment environments as explained below.

\vspace{0.05in}\noindent\textbf{Transport mechanism.}  It supports four types of transport mechanisms: (i) local, (ii) RDMA, (iii) GDR, and (iv) TCP (ZeroMQ).  In ``local'' processing, a client processes data on a local GPU without offloading.  Hence it only incurs processing and inference latency, but no delay from data movement.  This presents a lower bound on achievable end-to-end latency.

\begin{table}[t]
\centering
\caption{DNN models used.}
\label{tab:models}
\resizebox{.95\columnwidth}{!}{%
\begin{tabular}{|c|c|c|c|c|c|}
\hline
 \textbf{Model} & \textbf{Task}  & \textbf{GFLOPS} & \textbf{Input Shape} & \textbf{Output Shape} \\ \hline
  \texttt{MobileNetV3} & Classification & 0.06 & 3$\times$224$\times$224 & 1$\times$1000\\
 \texttt{ResNet50} & Classification & 4.1 & 3$\times$224$\times$224 & 1$\times$1000\\
  \texttt{EfficientNetB0} & Classification & 0.39 & 3$\times$224$\times$224 & 1$\times$1000\\
 \texttt{WideResNet101} & Classification & 22.81 & 3$\times$224$\times$224 & 1$\times$1000\\
  \texttt{YoloV4} & Detection & 128.46 & 3$\times$416$\times$416 & \begin{tabular}{@{}c@{}}S$\times$S$\times$3$\times$85, \\ S = $\{13, 26, 52\}$\end{tabular} \\
 \texttt{DeepLabV3\_ResNet50} & Segmentation & 178.72 & 3$\times$520$\times$520 & 2$\times$21$\times$520$\times$520    \\
 \hline
\end{tabular}%
}
\vspace{-2ex}
\end{table}

\begin{figure}[t]
\centering
    \begin{subfigure}{0.3\textwidth}
        \centering
       \includegraphics[width=\textwidth]{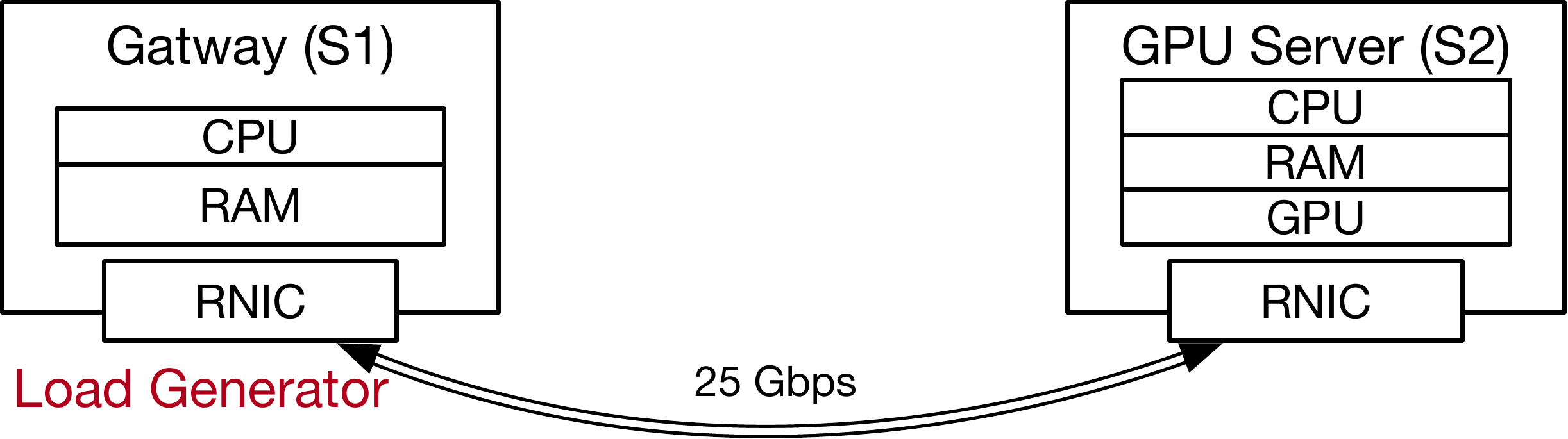}
       \caption{Direct connection}
       \label{fig:connection_direct_setup}
    \end{subfigure}
\\
    \begin{subfigure}{0.4\textwidth}
        \centering
        \includegraphics[width=\textwidth]{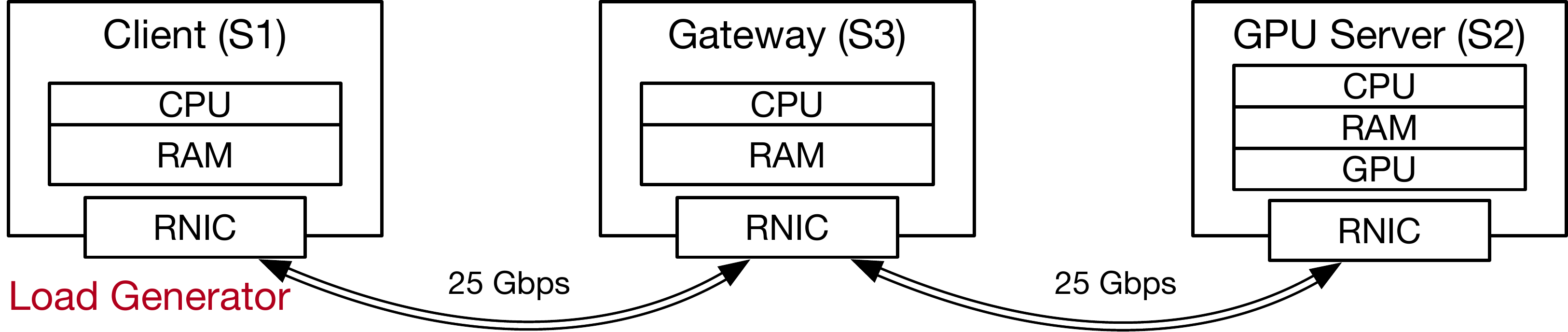}
        \caption{Proxied connection}
        \label{fig:connection_multilevel_setup}
       \vspace{-0.1cm}
    \end{subfigure}
\caption{Connection modes.}%
\vspace{-4ex}
\label{fig:connections_setup}
\end{figure}
\noindent\textbf{Connection mode.}
It supports two common connection scenarios between a client and a GPU server: (i) direct connection, and (ii) proxied connection.
They are compared in Figure~\ref{fig:connections_setup}. 
The direct connection mode illustrates the connectivity between a gateway and a GPU server within an edge computing facility. In this case, we deploy our load generator on the gateway server. Note that both the gateway and the GPU server must be equipped with RNICs to support hardware-accelerated transport.
The proxied connection mode represents the case where client-to-server communication is proxied by an intermediate gateway. To focus on the effect of networking rather than the gateway's scheduling decision, the gateway is configured to forward client requests to a fixed server. 

\noindent\textbf{Workload.} To evaluate the effect of different jobs and data sizes, we deploy several different DNN models as shown in Table~\ref{tab:models}.
The models differ in terms of functionality, model complexity, and communication overhead (i.e., input/output sizes). The classification models are trained with ImageNet~\cite{imagenet}, while detection and segmentation models are trained with Microsoft COCO~\cite{coco}.

\noindent\textbf{GPU configuration.} Finally, we evaluate common methods to control processing latency within a GPU by varying client concurrency, client priority, and GPU sharing modes, namely \texttt{multi-stream}, \texttt{multi-context}, and \texttt{MPS}.

\begin{table}[t]
\centering
\caption{Testbed configuration.}
\label{tab:server}
\resizebox{\columnwidth}{!}{%
\begin{tabular}{|c|c|c|c|c|}
\hline
 \textbf{Name}& \textbf{Server type} & \textbf{CPU} & \textbf{GPU} & \textbf{NIC}\\ \hline
 S1 & Dell PowerEdge R740 & Intel Xeon-G 6240 & - & ConnectX-5 25GbE \\ 
 S2 & Dell PowerEdge R740 & Intel Xeon-G 6240 & NVIDIA A2 & ConnectX-5 25GbE \\ 
 S3 & Dell PowerEdge R750 & Intel Xeon-G 6330 & - & ConnectX-5 25GbE \\ 
 \hline
\end{tabular}%
}
\vspace{-3ex}
\end{table}

\subsection{Implementation and Deployment}

In implementing the aforementioned system, we use NVIDIA OFED v5.6 for RDMA communication, and ZeroMQ v2.1 for TCP-based communication. For model-serving pipelines, we use CUDA toolkit v11.6.2, OpenCV v4.5.5, and TensorRT v8.4. The client and server are written in C++ and comprise $\sim$4.5k SLOC.
The model-serving system is deployed on three servers described in Table~\ref{tab:server}.
S2 is equipped with NVIDIA A2 GPU, which has 10 execution engines, 16 GB memory, and two copy-engines. All servers are running on Ubuntu v20.04 LTS and kernel v5.15.

In the rest of the paper, we utilize our platform to systematically examine the latency performance of model-serving under different scenarios. Starting from employing a single client session to identify the bottlenecks of the model-serving pipeline without resource sharing and contention, we further study the impact of concurrency when offload resources are subject to competition from multiple clients.
Lastly, we explore the trade-offs of the mechanisms used to tame the overhead of concurrency. In all cases, different transport, workload, and connection mode combinations are explored. 

\section{Single Client Performance}\label{sec:single}

\begin{figure}
\centering
    \begin{subfigure}{0.24\textwidth}
        \centering
       \includegraphics[width=\textwidth]{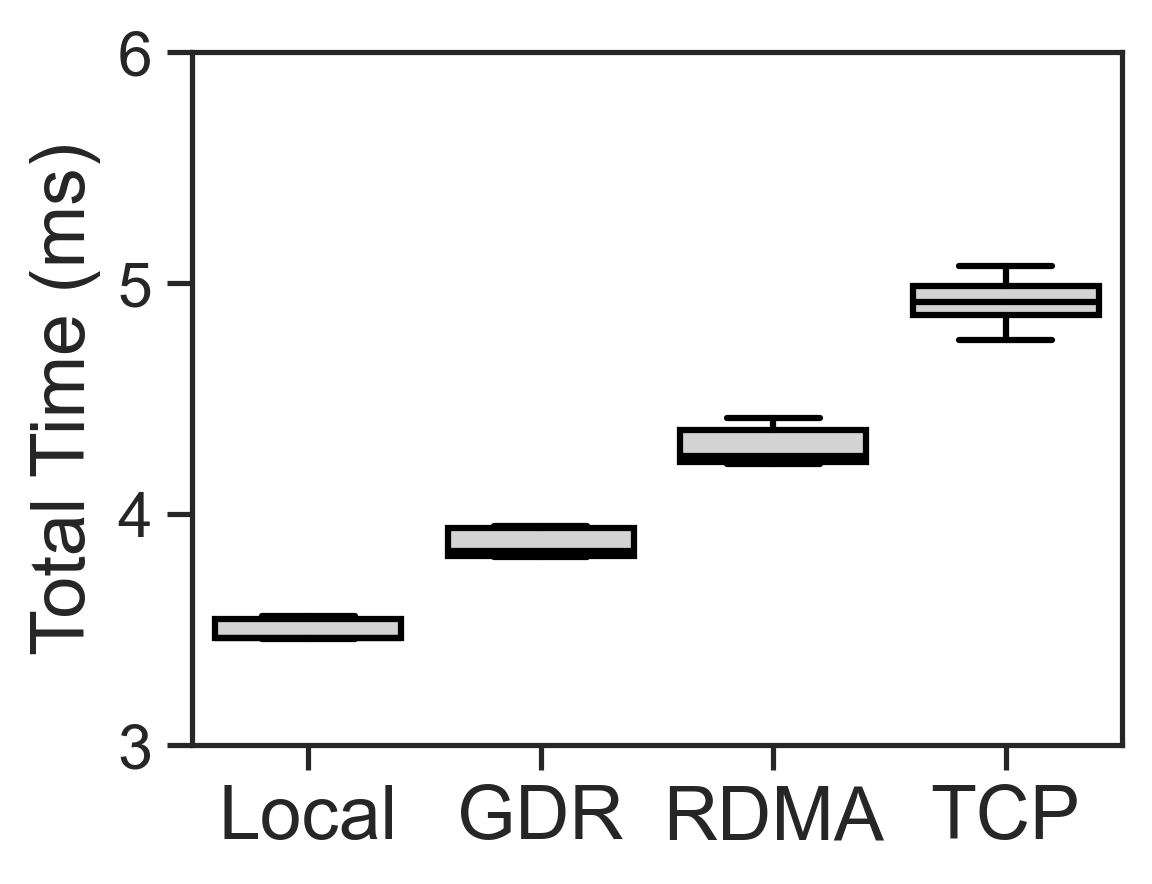}
       \caption{Raw images}
       \label{fig:single_b2b_resnet50_mat_total}
    \end{subfigure}
\hfill
    \begin{subfigure}{0.24\textwidth}
        \centering
        \includegraphics[width=\textwidth]{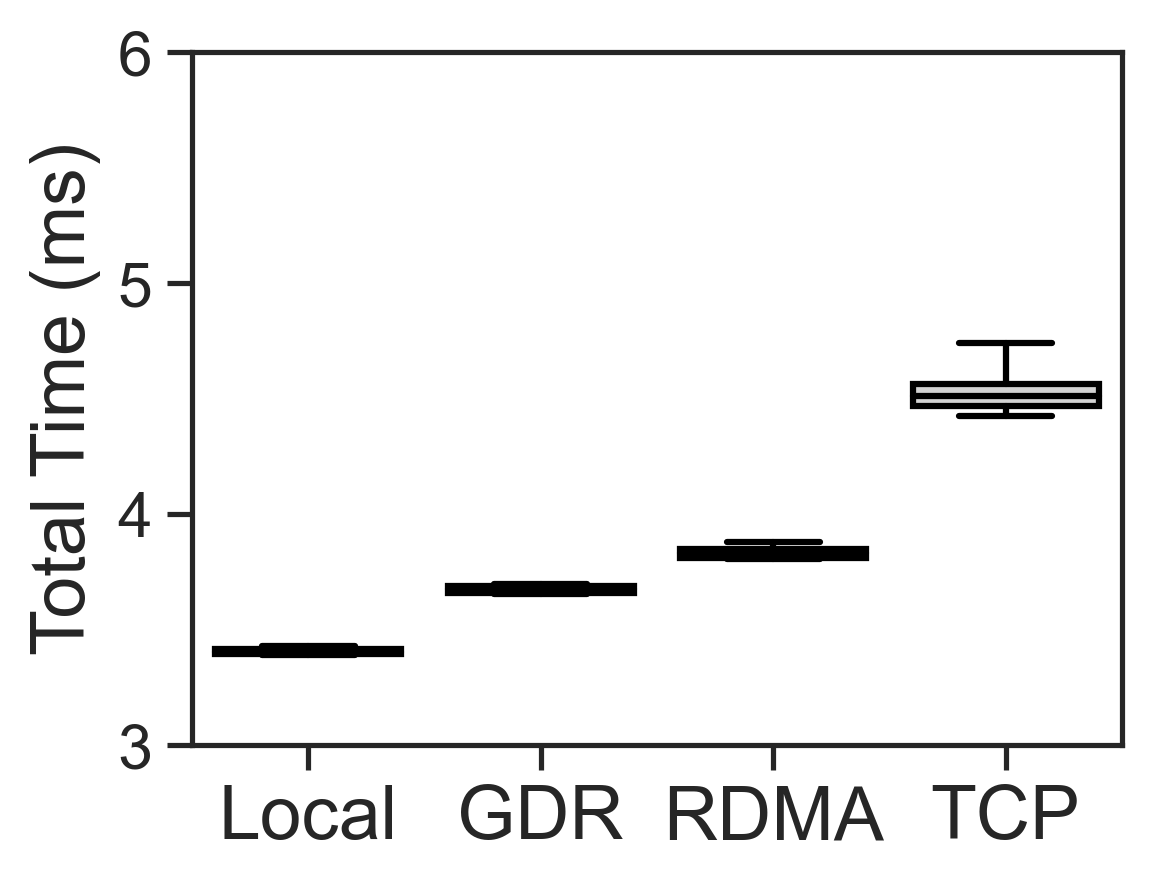}
        \caption{Preprocessed images}
        \label{fig:single_b2b_resnet50_processed_total}
    \end{subfigure}
\vspace{-4ex}
\caption{Total latency across mechanisms for \texttt{ResNet50}.}%
\label{fig:single_b2b_resnet50_total}
\vspace{-3ex}
\end{figure}

\begin{figure}
\centering
\begin{subfigure}{0.48\textwidth}
        \raggedright
       \includegraphics[width=\textwidth]{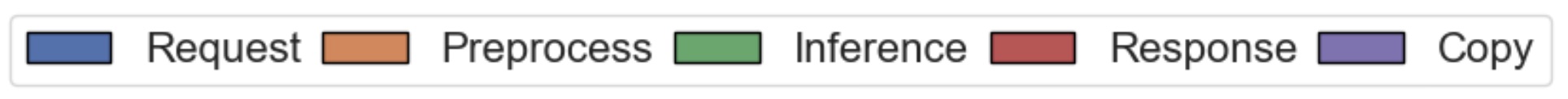}
    \end{subfigure}
    \\
    \begin{subfigure}{0.24\textwidth}
        \centering
       \includegraphics[width=\textwidth]{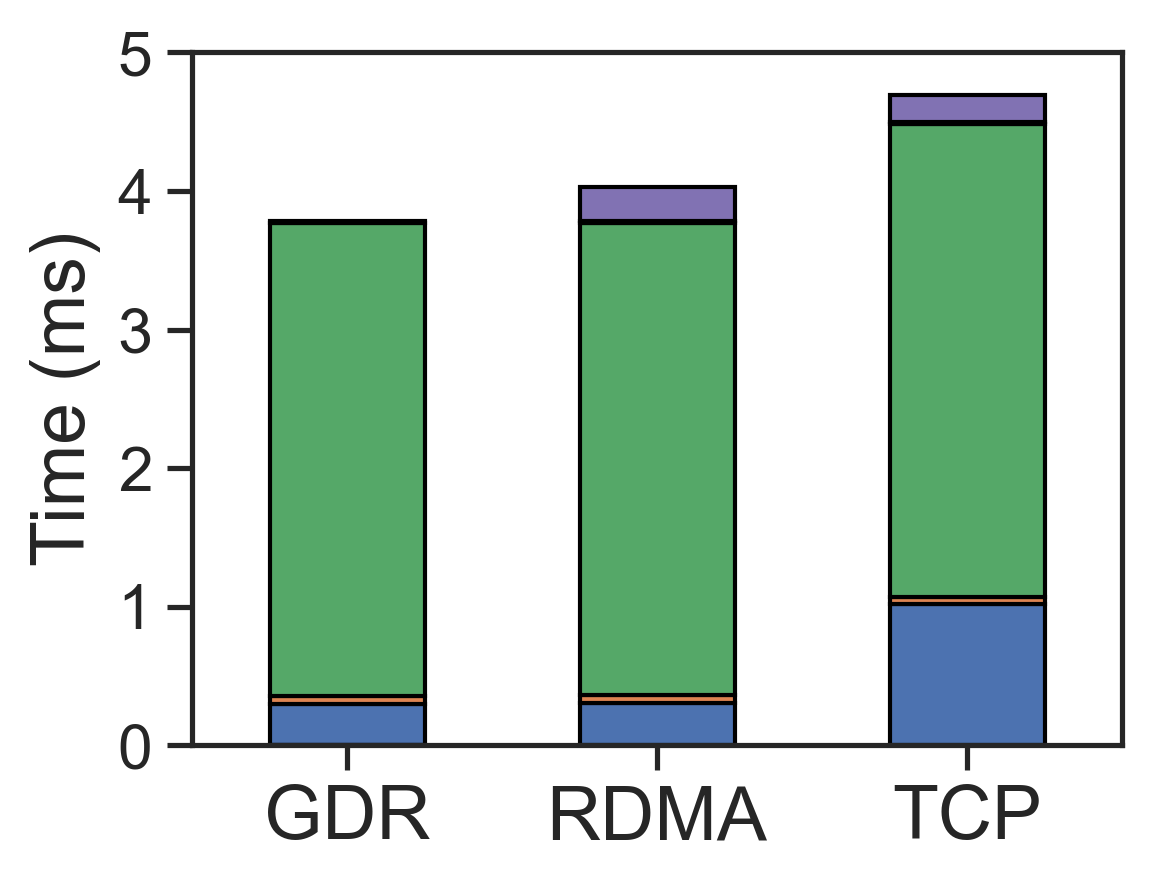}
       \caption{Raw images}
       \label{fig:single_b2b_resnet50_mat_breakdown}
    \end{subfigure}
\hfill
    \begin{subfigure}{0.24\textwidth}
        \centering
        \includegraphics[width=\textwidth]{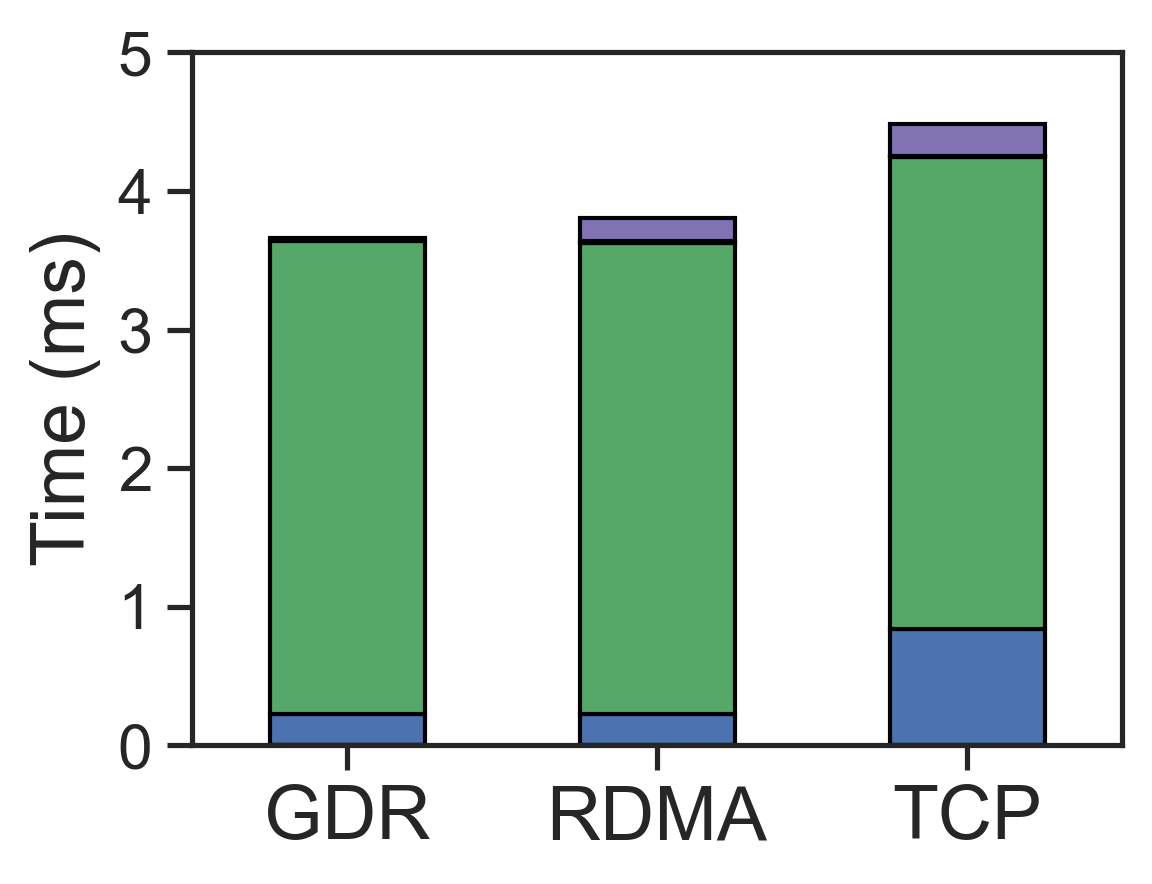}
        \caption{Preprocessed images}
        
        \label{fig:single_b2b_resnet50_processed_breakdown}
    \end{subfigure}%
\vspace{-1.5ex}
\caption{Latency breakdown across mechanisms for \texttt{ResNet50}.}%
\label{fig:single_b2b_resnet50_breakdown}
\vspace{-4ex}
\end{figure}

\begin{figure*}[t]
\centering
    \begin{subfigure}{0.4\linewidth}
        \centering
       \includegraphics[width=\textwidth]{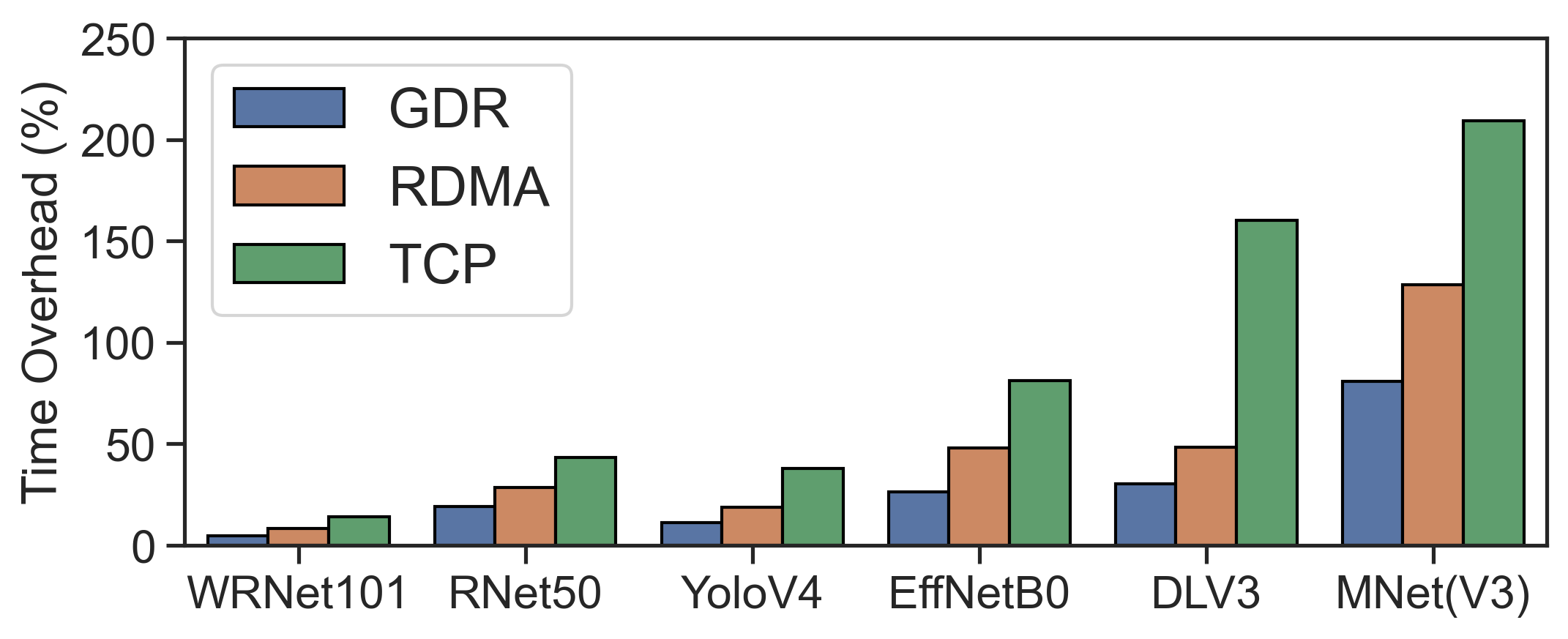}
       \caption{Raw Images}
       \label{fig:single_b2b_models_mat_overhead}
    \end{subfigure}
\hspace{1 cm}
    \begin{subfigure}{0.4\linewidth}
        \centering
        \includegraphics[width=\textwidth]{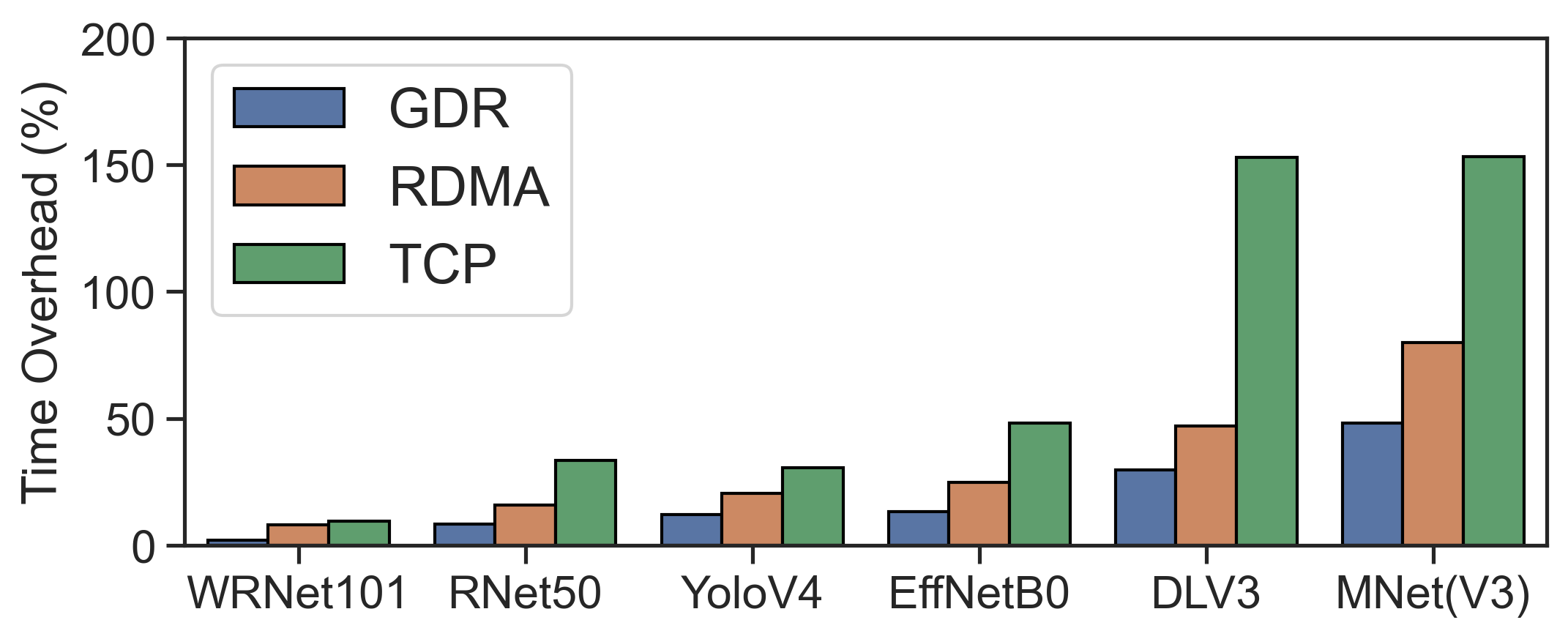}
        \caption{Preprocessed Images}
        \label{fig:single_b2b_models_processed_overhead}
    \end{subfigure}
\vspace{-1.5ex}
\caption{Model-serving latency overhead with respect to local processing for different DNN models.}%
\label{fig:single_b2b_models_overhead}
\vspace{-2ex}
\end{figure*}

\begin{figure*}[t]
\centering
\begin{subfigure}{0.60\textwidth}
        \centering
       \includegraphics[width=\textwidth]{figures/bd_legend.jpg}
    \end{subfigure}
    \\
    \begin{subfigure}{0.32\linewidth}
        \centering
        \includegraphics[width=\textwidth]{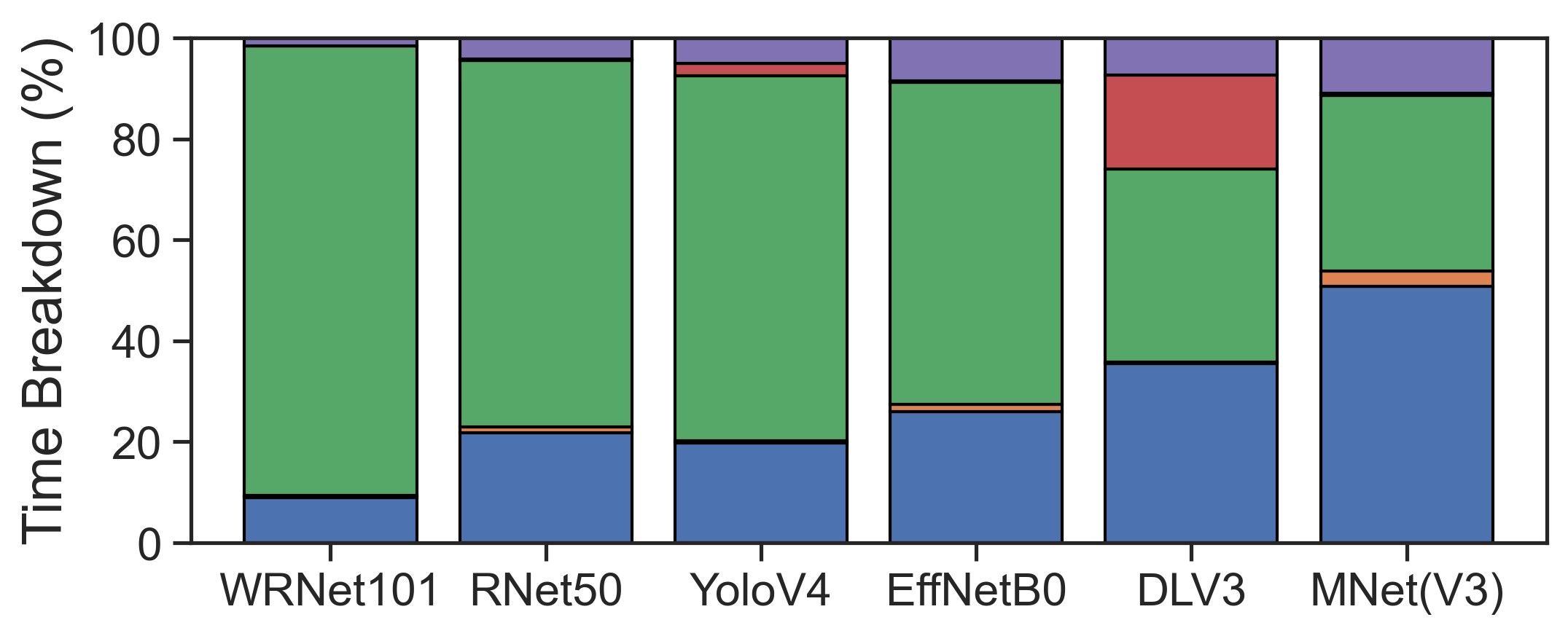}
        \caption{TCP}
    \label{fig:single_b2b_models_raw_bd_zmq}
    \end{subfigure}
\hfill
    \begin{subfigure}{0.32\linewidth}
        \centering
       \includegraphics[width=\textwidth]{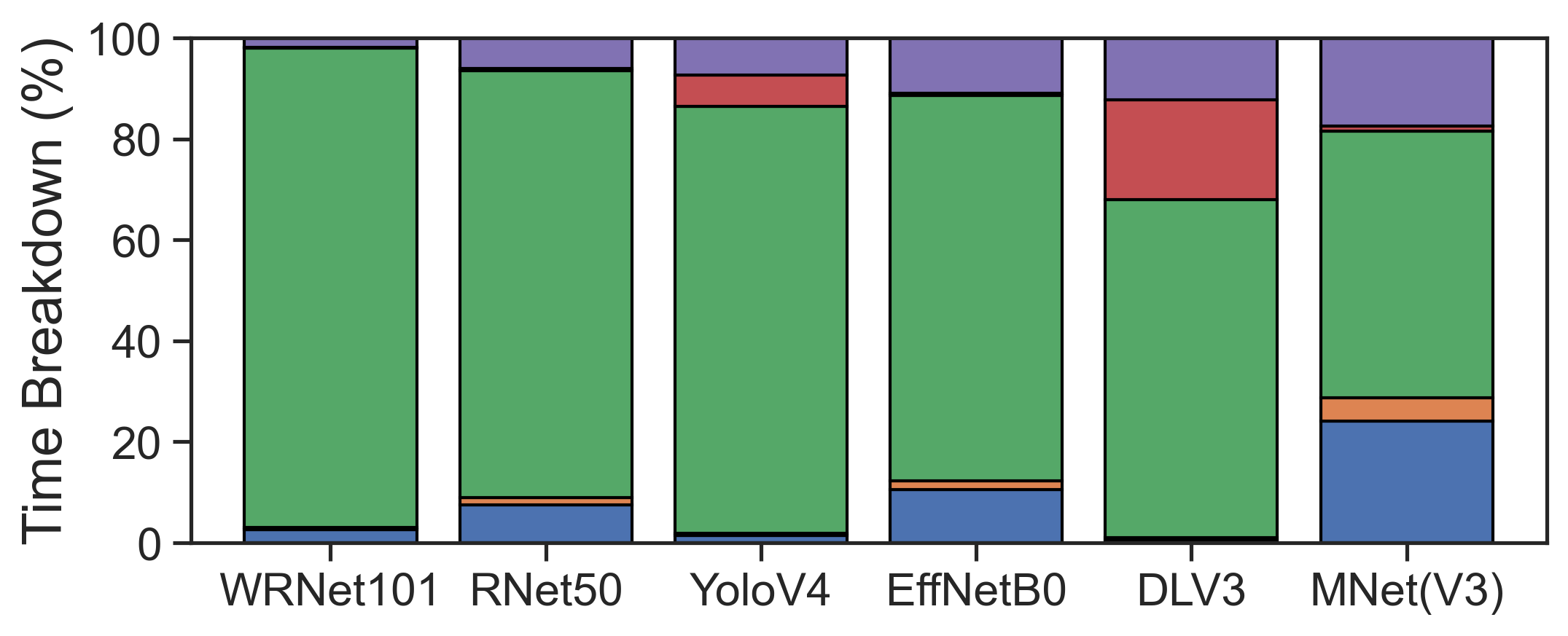}
       \caption{RDMA}
       \label{fig:single_b2b_models_raw_bd_rdma}
    \end{subfigure}
\hfill
    \begin{subfigure}{0.32\linewidth}
        \centering
        \includegraphics[width=\textwidth]{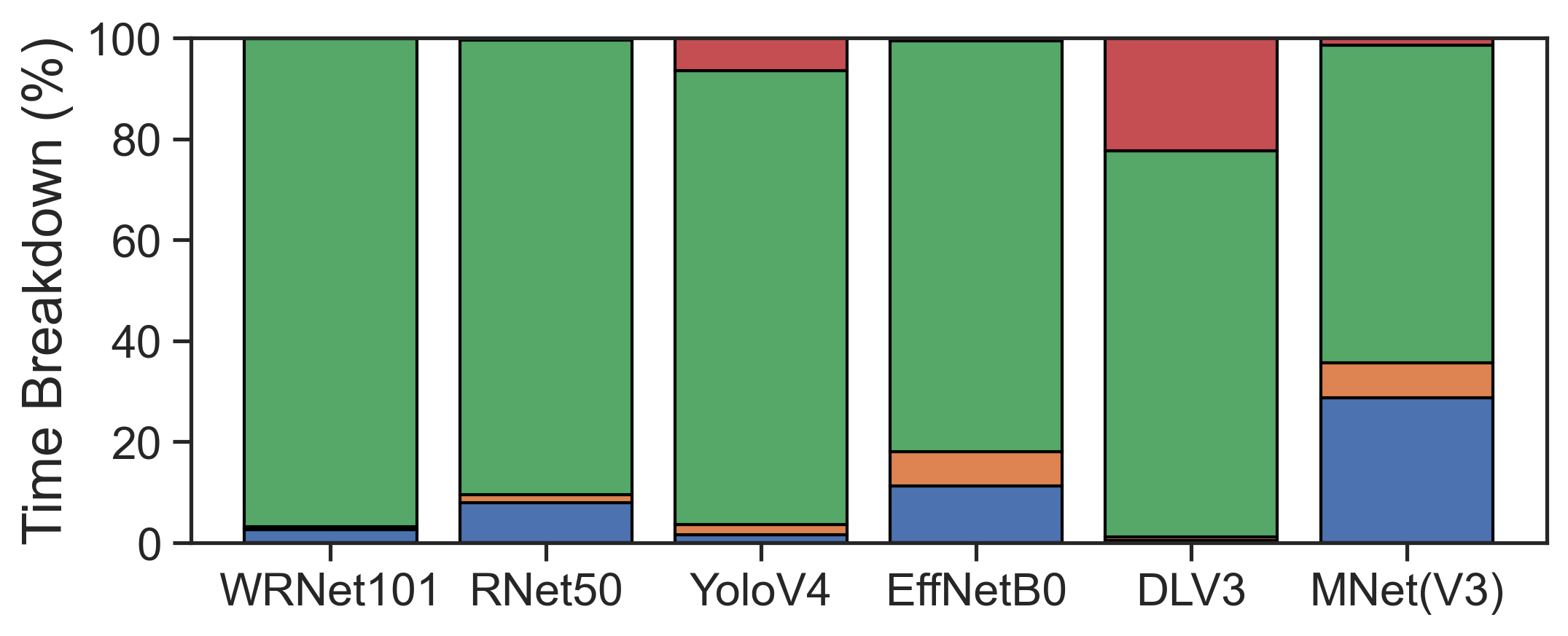}
        \caption{GPUDirect RDMA (GDR)}
        \label{fig:single_b2b_models_raw_bd_gdr}
    \end{subfigure}
\vspace{-1.5ex}
\caption{Latency breakdown across different transport mechanisms for different DNN models.}%
\vspace{-2ex}
\label{fig:single_b2b_models_bd}
\end{figure*}

In general, when a client offloads a model-serving computation to a server, its end-to-end latency is determined by how the client's model-serving request is delivered to the target compute resource (transport delay), as well as how the request is executed within the compute resource (execution delay).  Within an edge computing facility, the transport delay can vary with different transport mechanisms, while the execution delay can be influenced by how the compute resource is shared to handle  concurrent client requests.  In the first set of evaluations, we focus on the transport delay, while discounting the effect of a specific GPU scheduling algorithm.  For this purpose, we evaluate the latency performance of model-serving across different transport mechanisms when running a \emph{single client}, which shows the performance of model-serving without any interference from sharing edge network or compute resources.

\subsection{Direct Connection}\label{sec:single_direct}
We start by evaluating model-serving performance in the most simplistic scenario, where model-serving requests are handled in the direct connection mode. In this mode, we exclude the the client-to-edge latency and focus on the latency with the edge network fabric. In this case, we run the load generator on the gateway server itself.
In Figure~\ref{fig:single_b2b_resnet50_total}, we compare model-serving latency across different transport mechanisms when \texttt{ResNet50} is used (Table~\ref{tab:models}). We add ``local processing'' as a reference. We repeat the experiments with and without preprocessing.  
The figure shows that GDR and RDMA perform better than TCP-based transport. When the server performs preprocessing (Figure~\ref{fig:single_b2b_resnet50_total}(a)), GDR and RDMA incur 20.3\% and 11.4\% less latency than TCP, respectively.  Without preprocessing (Figure~\ref{fig:single_b2b_resnet50_total}(b)), GDR and RDMA lead to 23.2\% and 15.2\% shorter delay than TCP, respectively.  The relative performance of hardware-accelerated transports, compared to TCP, is more pronounced when preprocessing is not needed because the overall model-serving pipeline takes less time to execute.  Compared to local processing, GDR-based model-serving adds as low as \SIrange[range-phrase=--,range-units=single]{0.27}{0.53}{\ms}, while TCP adds \SIrange[range-phrase=--,range-units=single]{1.2}{1.5}{\ms}, depending on whether or not preprocessing is needed.

To understand the source of difference, we plot the latency breakdown in Figure~\ref{fig:single_b2b_resnet50_breakdown}.
The figure shows that the difference between GDR/RDMA and TCP comes from data transfer time. For example, TCP-based transport takes \SI{0.73}{\ms} and \SI{0.61}{\ms} more to send raw and preprocessed data than GDR and RDMA-based counterparts, respectively. GDR outperforms RDMA as it skips H2D and D2H copies, which saves extra \SI{0.3}{\ms} and \SI{0.2}{\ms} when handling raw and preprocessed data, respectively. This highlights the advantage of GDR and quantifies potential bottlenecks across protocols when the number of clients increases, which will be demonstrated in Section~\ref{sec:scalability}.

To generalize these findings, we repeat the experiments with other ML models of varied complexity and I/O sizes, as listed in Table~\ref{tab:models}. 
Figure~\ref{fig:single_b2b_models_overhead} shows the latency overhead with respect to local processing for different models.  That is, with each model, it shows how much longer latency is incurred by offloading the model compared to executing it locally.  Figures~\ref{fig:single_b2b_models_overhead}(a) and (b) show the latency overhead when sending raw and preprocessed images, respectively. In both cases, GDR outperforms the alternatives as expected. However, the overhead greatly varies across models. It shows that smaller models tend to have higher overhead than bigger models, and that models with larger I/O have higher overhead as well. This is because smaller models and models with larger I/O sizes have a higher fraction of time spent in the communication stage, making the role of transport mechanisms more vital than bigger models with smaller I/O sizes.
For example, offloading \texttt{MobileNetV3} adds at least 80.8\% and 48.1\% overhead compared to local processing, while offloading \texttt{WideResNet101} adds just about 4.5\% and 2\% overhead when serving raw and preprocessed images, respectively. Models with large I/O sizes (e.g., \texttt{DeepLabV3}) also show very high overhead, especially with TCP.

To quantify these overheads, Figure~\ref{fig:single_b2b_models_bd} depicts the fraction of time spent in each stage.  The result confirms our hypothesis. For instance, when serving \texttt{MobileNetV3}, 62\%, 42\%, and 30\% of total time is spent in data movement (\texttt{copy-time} + \texttt{request-time} + \texttt{response-time}) when serving requests over TCP, RDMA, and GDR, respectively, while in \texttt{WideResNet101}, the fraction of communication overhead does not surpass 10\% in all cases. The result also shows the merits of hardware-accelerated transport for large I/O.
For example, when serving raw data using \texttt{DeepLabV3}, TCP spends 60\% of the overall latency in data movement, while RDMA and GDR spend only 32\% and 23\%. In this case, the overhead of large I/O size is translated to higher latency difference, where TCP-based transport adds \SI{71}{\ms} and \SI{68}{\ms}, compared to GDR and RDMA-based transports, respectively.
Note that, with more powerful accelerators and more I/O-intensive immersive application offloading, the fraction of time spent in actual processing will become smaller, which will further increase the importance of transport methods.

\begin{figure}[t]
    \vspace{-1ex}
    \centering
    \includegraphics[width=0.4\textwidth]{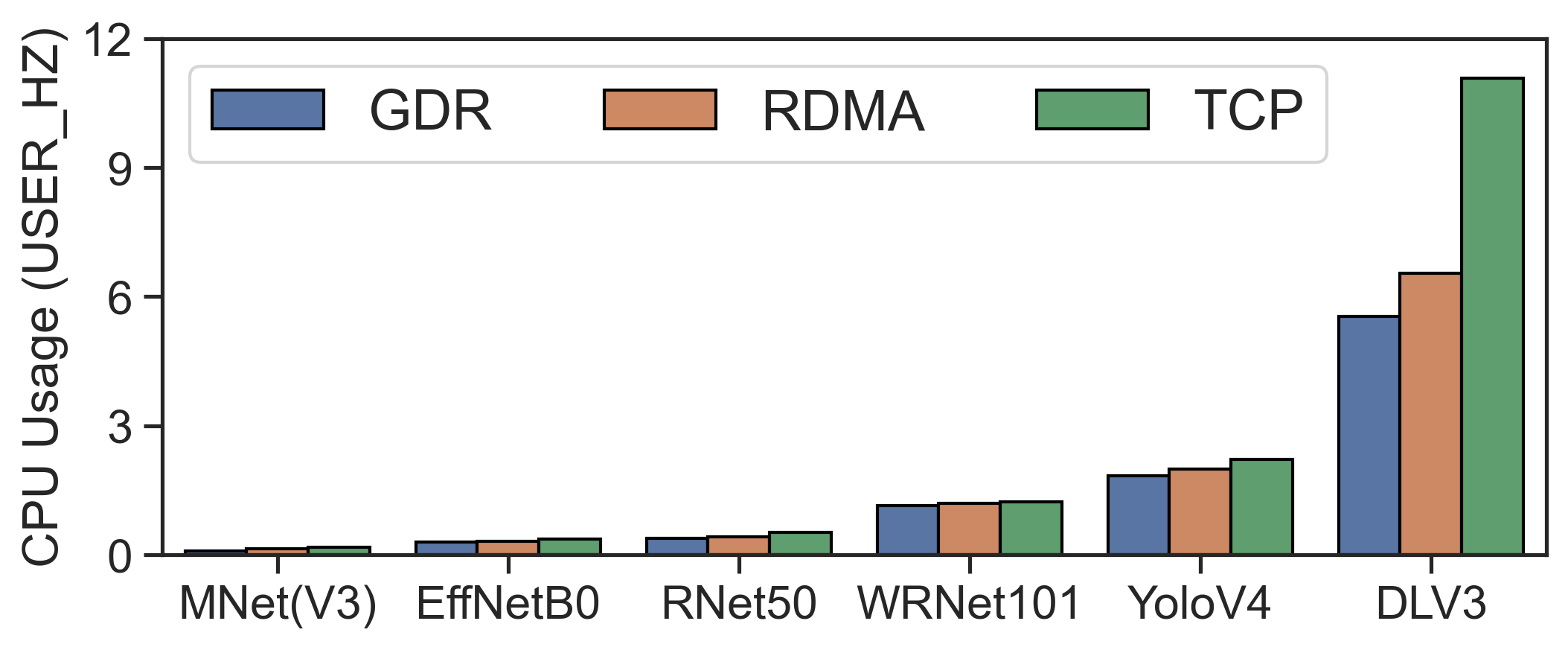}
    \vspace{-2ex}
    \caption{CPU Usage across models.}
    \label{fig:models_cpu-usage}
    \vspace{-4ex}
\end{figure}

Finally, Figure~\ref{fig:models_cpu-usage} shows the CPU usage per request across different models. 
It shows that TCP-based transport incurs the highest CPU usage as the CPU is involved in communication. The overhead is most visible when serving \texttt{DeepLabV3} as its I/O size is high, where TCP adds 100\% more CPU usage than GDR-based transport. It also shows that issuing copy operations for RDMA adds only a minor effect.

\begin{figure}[t]
    \centering
    \includegraphics[width=0.4\textwidth]{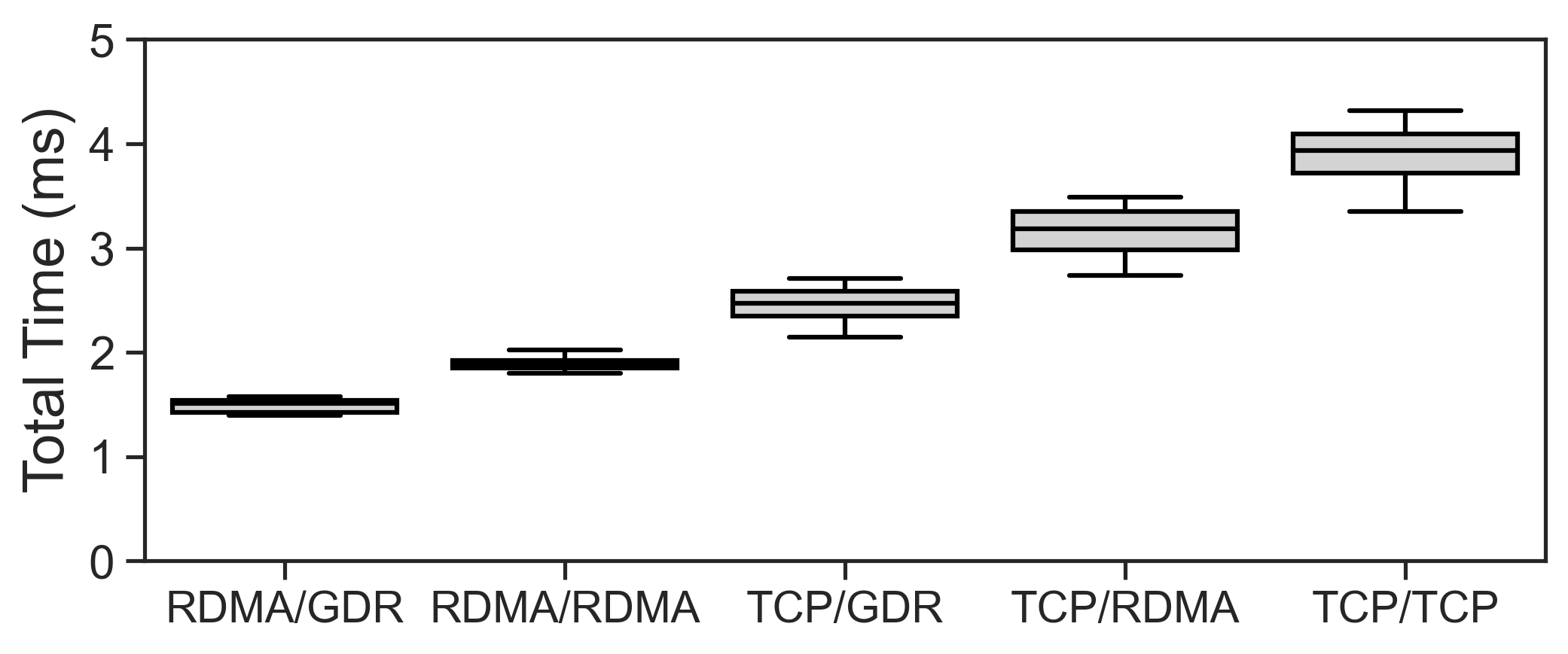}
    \vspace{-2ex}
    \caption{End-to-end latency with proxied connection.}
    \label{fig:single_proxy}
    \vspace{-3ex}
\end{figure}

\begin{figure*}
\centering
    \begin{subfigure}{0.4\linewidth}
       	\centering
	   \includegraphics[width=\textwidth]{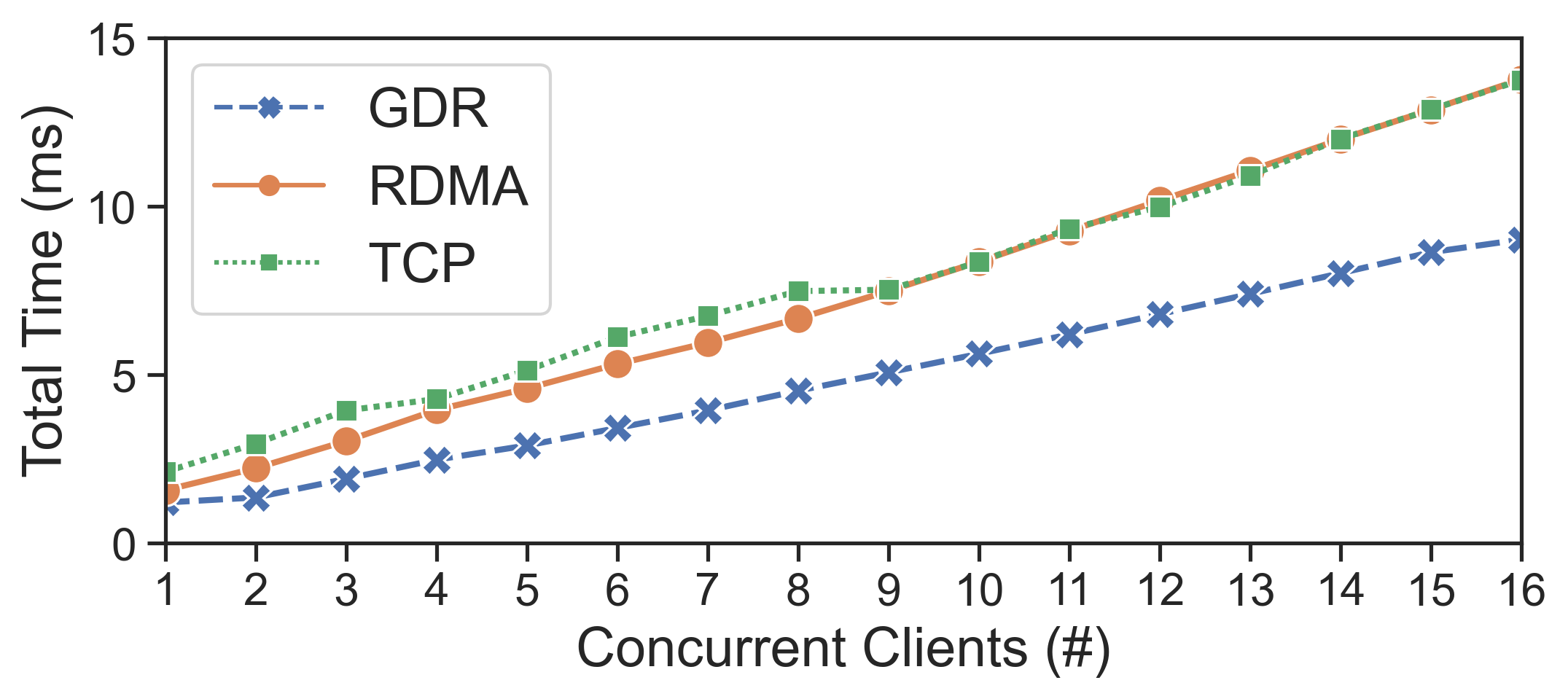}
        \vspace{-3ex}
	   \caption{\texttt{MobileNetV3}}
	   \label{fig:multiple_b2b_mobilenet_latency}
    \end{subfigure}
\hspace{1cm}
    \begin{subfigure}{0.4\linewidth}
        \centering
        \includegraphics[width=\textwidth]{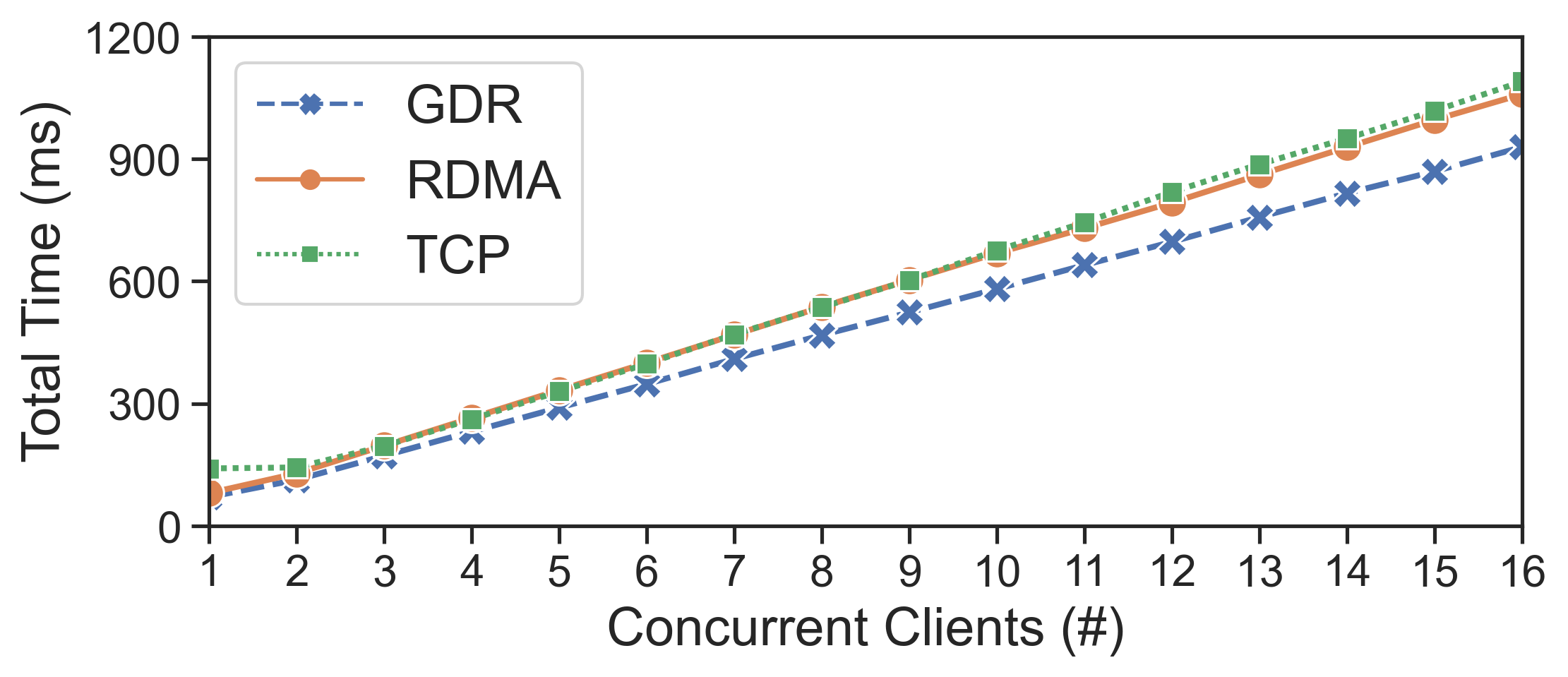}
        \vspace{-3ex}
        \caption{\texttt{DeepLabV3}}
        \label{fig:multiple_b2b_deeplab_latency}
    \end{subfigure}
\caption{Total time across clients when processing raw images.}%
\vspace{-3ex}
\label{fig:multiple_b2b_latency}
\end{figure*}

\subsection{Proxied Connection}\label{sec:single-ml}
Next, we switch to a more realistic scenario, where a client sends its requests to an intermediate gateway which then steers the requests to an appropriate GPU server.
To focus on the effect of transport mechanisms in such ``proxied-connection'' scenario, we exclude the overhead of server selection within the gateway by letting the gateway forward requests to a fixed server.
In this scenario, client-to-gateway and gateway-to-server communication can be realized with available transport mechanisms independently.  This results in the following configuration pairs for client-to-gateway and gateway-to-server transports: (i) RDMA/GDR, (ii) RDMA/RDMA, (iii) TCP/GDR, and (iv) TCP/RDMA. Lastly, we add TCP/TCP as a representative of existing model-serving frameworks.

Figure \ref{fig:single_proxy} plots model-serving latency results when a client submits raw data to \texttt{MobileNetV3}.
It shows that hardware-accelerated transports can improve model-serving latency even if applied only to the last hop of the communicating path.
Compared to end-to-end TCP connections (i.e., TCP/TCP), adopting hardware-accelerated transport between the gateway and the GPU server saves 23\% and 57\%, when replaced with RDMA (i.e., TCP/RDMA) and GDR (i.e., TCP/GDR). The results also show that TCP introduces higher performance variation, but the usage of hardware-accelerated communication, even partially, can reduce its effects.

\noindent\emph{\textbf{Key takeaways:}
GDR can provide significant gains when communication comprises a high fraction of end-to-end latency. Hardware-accelerated transport can alleviate the overhead of proxied connections which are common in large-scale and dynamic model-serving environments.}

\section{Performance Scalability}\label{sec:scalability}
\begin{figure*}
\centering
\begin{subfigure}{0.6\textwidth}
        \raggedright
       \includegraphics[width=\textwidth]{figures/bd_legend.jpg}
    \end{subfigure}
    \\
    \begin{subfigure}{0.32\textwidth}
       	\centering
	   \includegraphics[width=\textwidth]{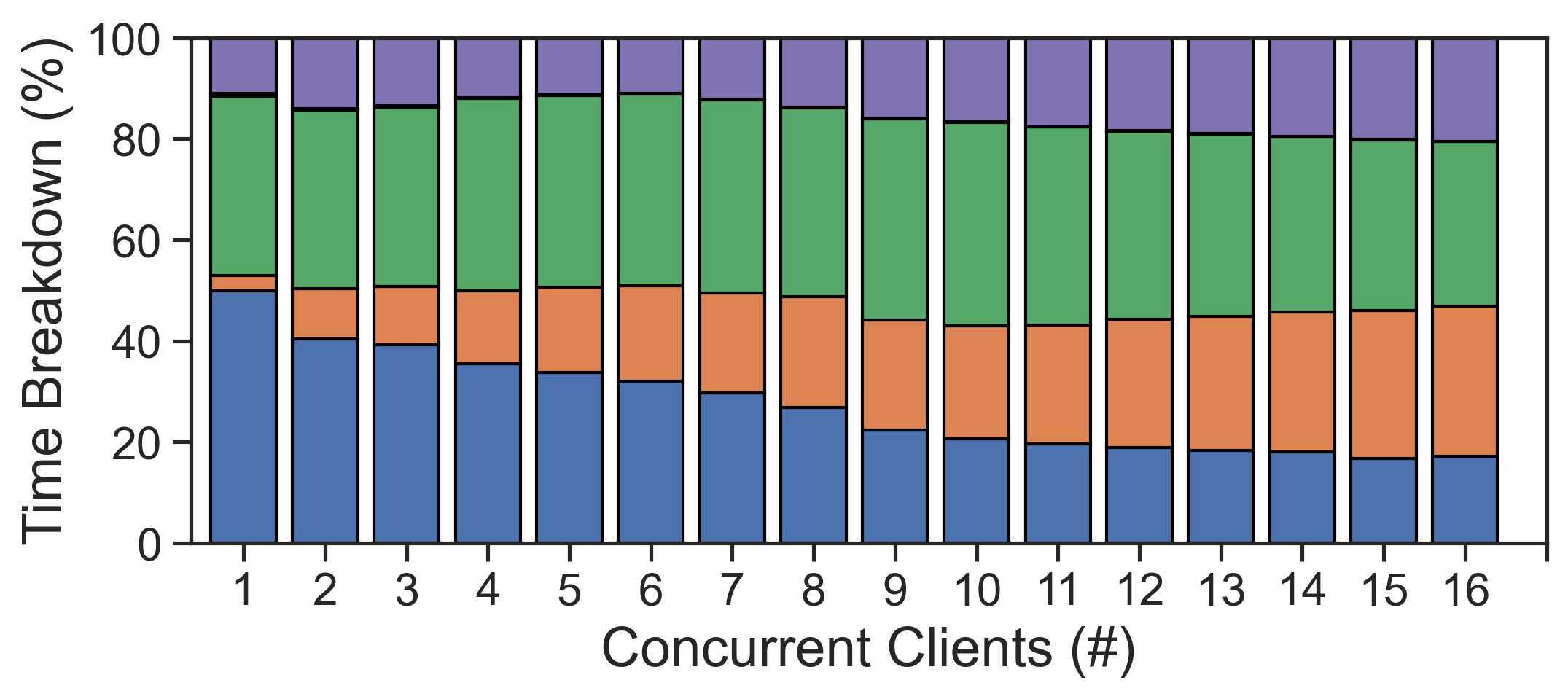}
	   \caption{TCP}
	   \label{fig:multiple_b2b_mat_mobilenetv3_overhead_zmq}
    \end{subfigure}
\hfill
    \begin{subfigure}{0.32\textwidth}
       	\centering
	   \includegraphics[width=\textwidth]{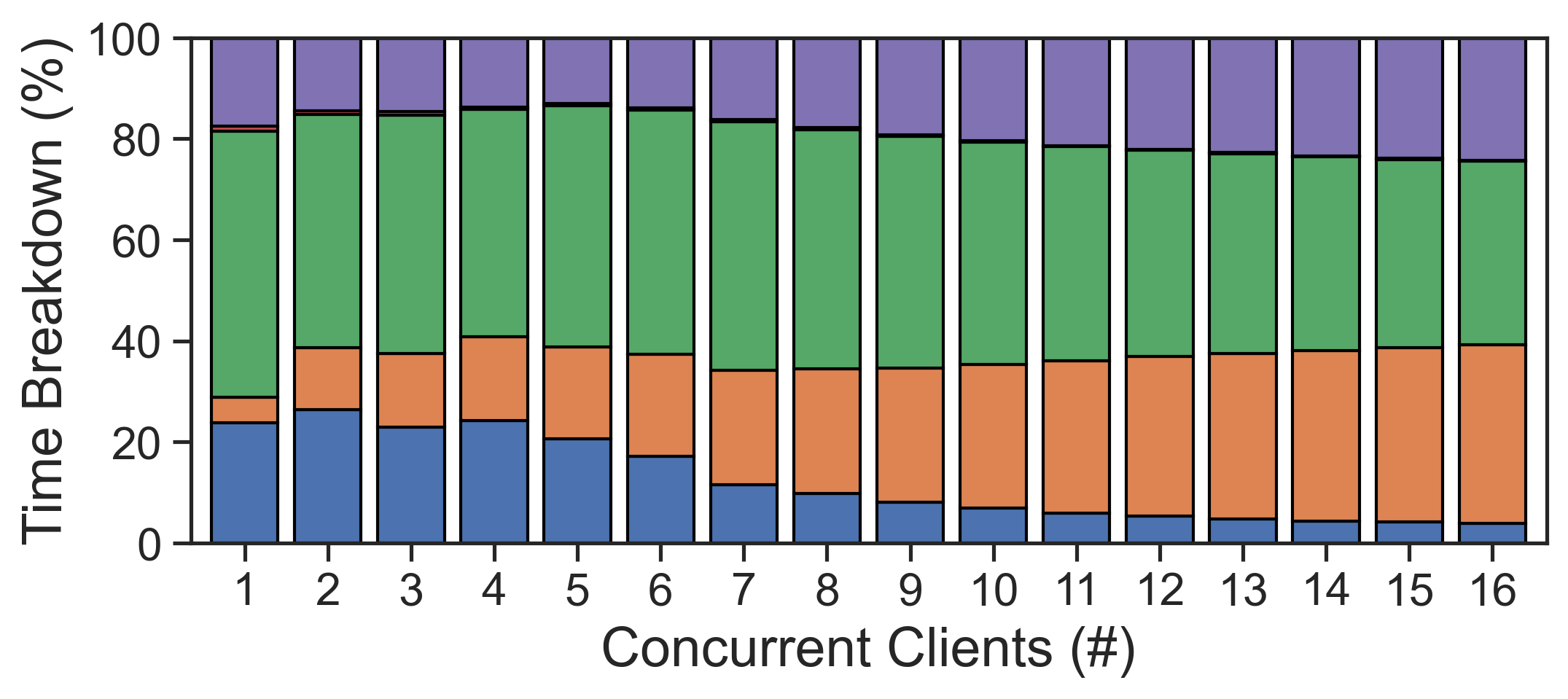}
	   \caption{RDMA}
	   \label{fig:multiple_b2b_mat_mobilenetv3_overhead_rdma}
    \end{subfigure}
\hfill
    \begin{subfigure}{0.32\textwidth}
        \centering
        \includegraphics[width=\textwidth]{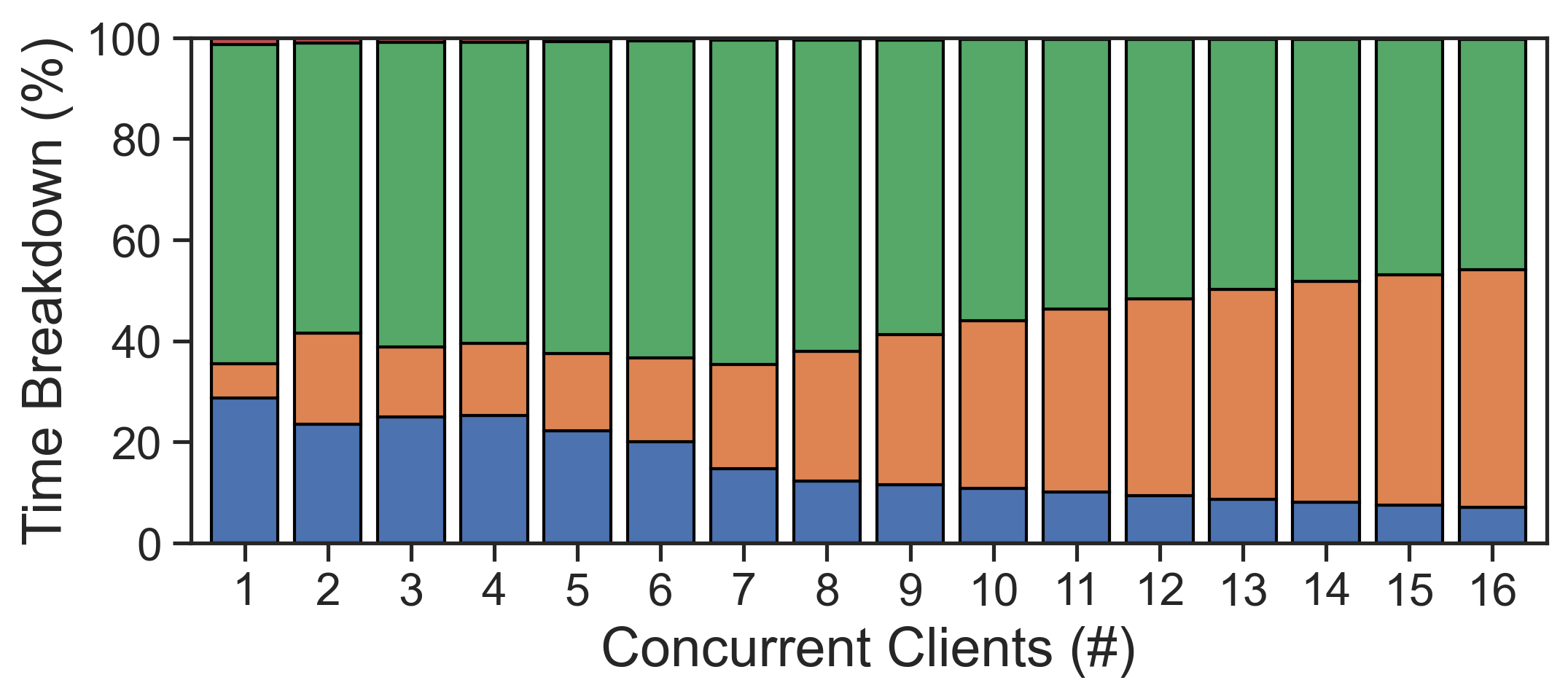}
        \caption{GDR}
        \label{fig:multiple_b2b_mat_mobilenetv3_overhead_GPUDirect}
    \end{subfigure}
\vspace{-1.5ex}
\caption{Latency breakdown when serving raw images using \texttt{MobileNetV3}.}%
\vspace{-2ex}
\label{fig:multiple_b2b_mat_mobilenetv3_overhead}
\end{figure*}

\begin{figure*}
\vspace{-2ex}
\centering
    \begin{subfigure}{0.32\textwidth}
       	\centering
	   \includegraphics[width=\textwidth]{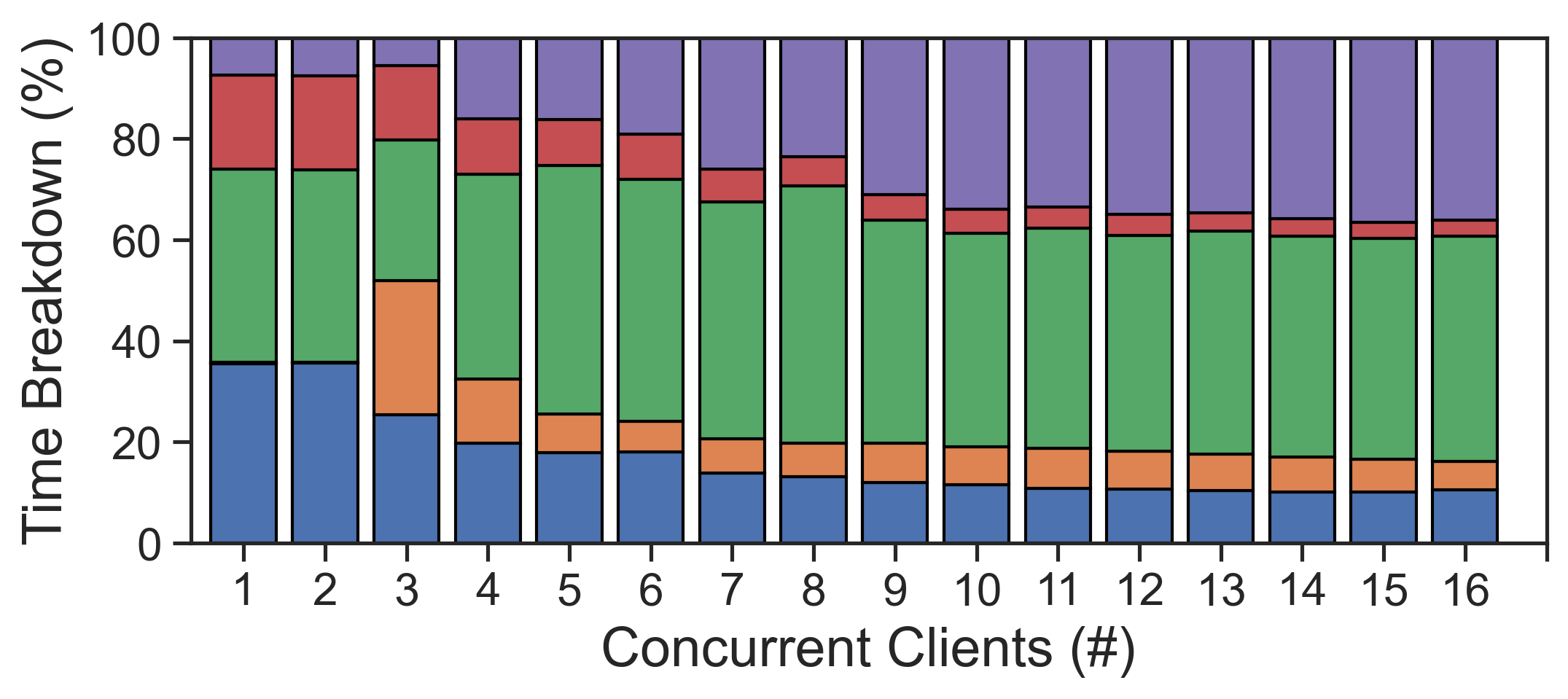}
	   \caption{TCP}%
\label{fig:multiple_b2b_mat_deeplabv3_overhead_zmq}
    \end{subfigure}
\hfill
    \begin{subfigure}{0.32\textwidth}
       	\centering
	   \includegraphics[width=\textwidth]{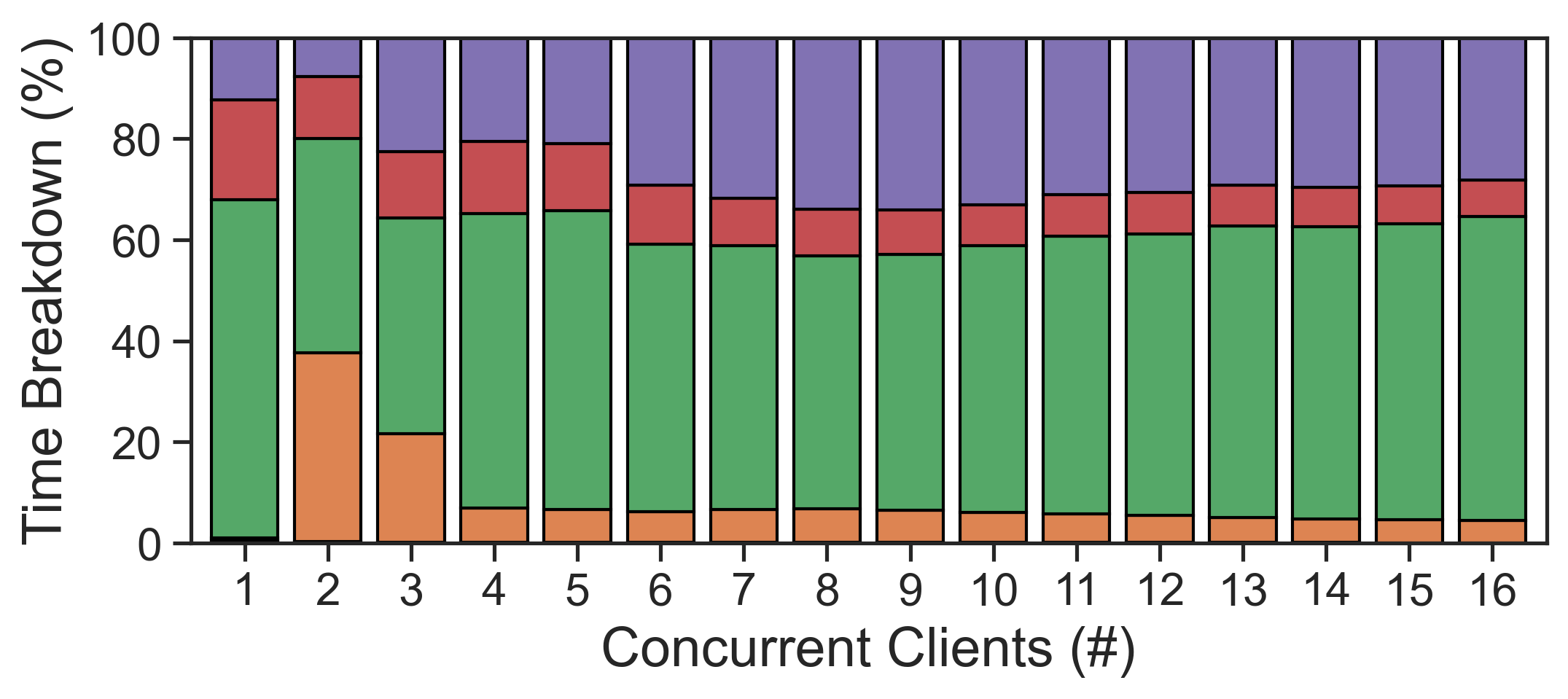}
	   \caption{RDMA}%
	   \label{fig:multiple_b2b_mat_deeplabv3_overhead_rdma}
    \end{subfigure}
\hfill
    \begin{subfigure}{0.32\textwidth}
        \centering
        \includegraphics[width=\textwidth]{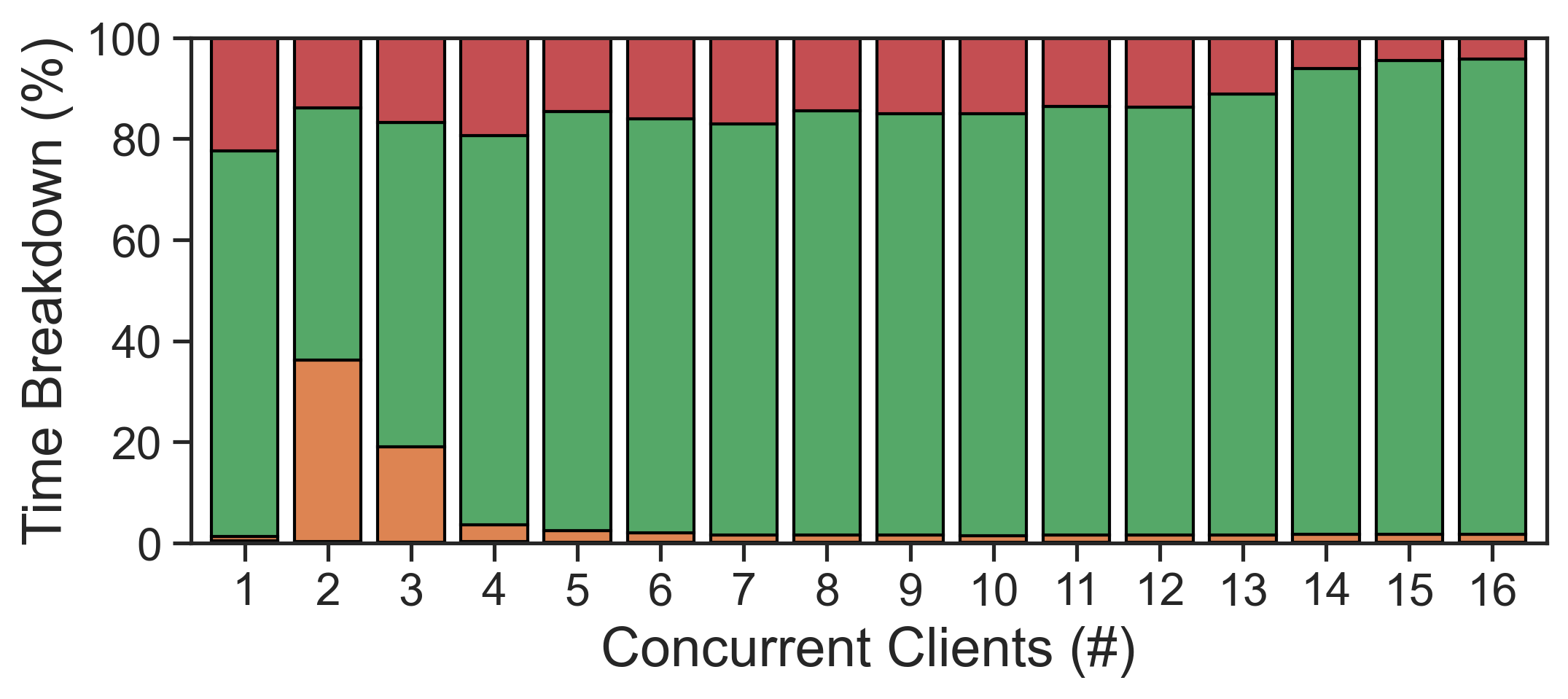}
        \caption{GDR}%
    \label{fig:multiple_b2b_mat_deeplabv3_overhead_gpudirect}
    \end{subfigure}
\vspace{-1.5ex}
\caption{Latency breakdown when serving raw images using \texttt{DeepLabV3}.}%
\vspace{-2ex}
\label{fig:multiple_b2b_mat_deeplabv3_overhead}
\end{figure*}

Next, we study performance scalability and the effect of sharing the compute infrastructure across multiple clients. In this set of experiments, our load generator starts multiple instances of a client application, each issuing model-serving requests concurrently. On the GPU-server side, requests from each client are handled by a dedicated stream. 

\subsection{Direct Connection}
First, we evaluate performance scalability using the direct connection mode. 
Figure \ref{fig:multiple_b2b_latency} shows the model-serving latency across a varying number of clients when \texttt{MobileNetV3} and \texttt{DeepLabV3} are used with raw images. In both cases, GDR outperforms both RDMA and TCP, and the gap between the two increases with the number of clients. For instance, with 16 clients, GDR saves \SI{4.7}{\ms} and \SI{160}{\ms} for \texttt{MobileNetV3} and \texttt{DeepLabV3}, respectively, compared to TCP. Surprisingly, however, the gain from using RDMA is lost with more clients, making its performance equivalent to that of TCP.

To understand the reason for this behavior, we examine how latency breakdown changes as we increase the number of clients. Figures \ref{fig:multiple_b2b_mat_mobilenetv3_overhead} and \ref{fig:multiple_b2b_mat_deeplabv3_overhead} show the fraction of time spent in each stage when serving raw images using \texttt{MobileNetV3} and \texttt{DeepLabV3}. As the number of clients changes, different models and transport mechanisms develop different bottlenecks. For instance, for \texttt{MobileNetV3}, the fraction of processing time (\texttt{preprocessing-time} + \texttt{inference-time})  increases from 38\% to 62\%, from 58\% to 72\%, and from 70\% to 92\% when TCP, RDMA, and GDR are used, respectively. Having the processing delay as the largest component makes network overheads negligible, which is a desired goal in edge offloading. 
However, for RDMA and TCP, increasing the copy time presents a steady source of overhead, where the GPU copy engine becomes the bottleneck. This explains the performance similarity between RDMA and TCP. The figure also shows that network I/O (request and response) never becomes a bottleneck in these two cases.  

On the other hand, when the more complex \texttt{DeepLabV3} is served, processing time has less impact on the performance, especially for RDMA and TCP. For instance, in case of GDR shown in Figure~\ref{fig:multiple_b2b_mat_deeplabv3_overhead}(c), processing time dominates the pipeline, which is indicated by the low transport overhead. However, in case of TCP and RDMA (Figures \ref{fig:multiple_b2b_mat_deeplabv3_overhead}(a) and \ref{fig:multiple_b2b_mat_deeplabv3_overhead}(b)), although processing time still dominates the overhead, the copy-time overhead becomes significant. The copy-time changes from 7\% to 36\% (\SIrange[range-phrase=--,range-units=single]{10}{366}{\ms}) for TCP, and from 12\% to 28\% (\SIrange[range-phrase=--,range-units=single]{9}{264}{\ms}) for RDMA. 

\subsection{Proxied Connection}
\begin{figure}
    \vspace{-1ex}
    \centering
    \includegraphics[width=0.4\textwidth]{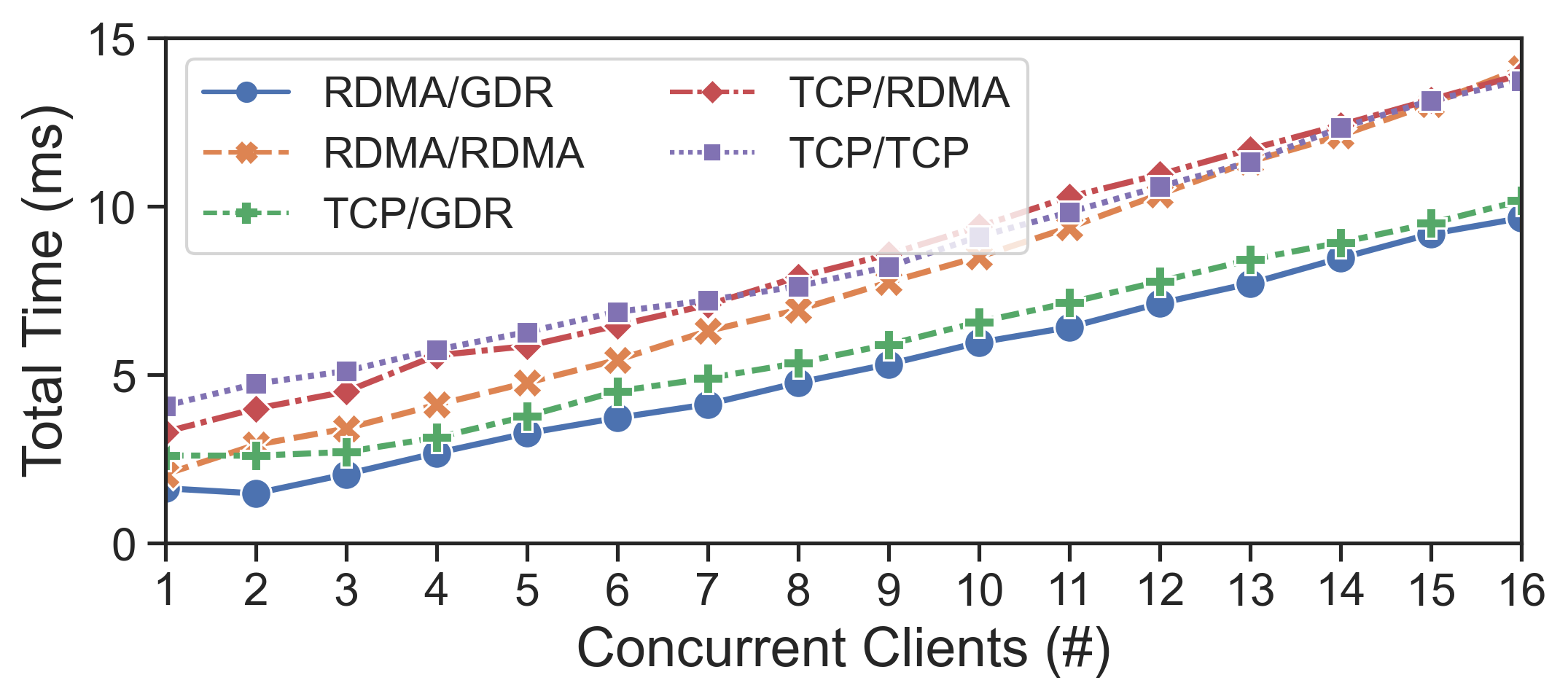}
    \vspace{-2ex}
    \caption{Performance scalability with proxied connection.}
    \vspace{-4ex}
    \label{fig:proxy_scalability}
\end{figure}
Figure \ref{fig:proxy_scalability} shows performance scalability in the proxied connection mode. Similar to Section~\ref{sec:single-ml}, we compare five different configurations for the proxied connection when serving raw images with \texttt{MobileNetV4}: (i) RDMA/GDR, (ii) RDMA/RDMA, (iii) TCP/GDR, (iv) TCP/RDMA, and (v) TCP/TCP. The figure shows that, as the number of clients increases, the behavior of different configurations changes in a counter-intuitive way, compared to the single client case discussed in Section \ref{sec:single-ml}. For instance, TCP/GDR outperforms the hardware-accelerated connection (RDMA/RDMA),  and it even becomes comparable to the RDMA/GDR case. Moreover, the performance of end-to-end TCP-based transport (TCP/TCP) becomes similar to that of hardware-accelerated transport in the last hop (TCP/RDMA) or end-to-end hardware-accelerated network (RDMA/RDMA). The resulting performance similarity between these methods (RDMA/RDMA, TCP/RDMA, and TCP/TCP) is caused by the bottleneck created by the copy engine, as explained earlier.
We note that the usage of GDR in the last hop saves 27\% compared to end-to-end TCP-based transport, while adding only 4\% compared to the best case (RDMA/GDR). 

\noindent\emph{\textbf{Key takeaways:}
The H2D/D2H copy quickly becomes a bottleneck in model-serving, which can easily remove any gain from hardware-accelerated RDMA communication. By skipping this step, GDR can yield higher scalability. This also applies to proxied communication, where skipping GPU's copy-engine with GDR greatly benefits the end-to-end latency.  Adopting GDR at the last hop communication is comparable to utilizing hardware-accelerated communication end-to-end in terms of the overall model-serving latency.}
\section{GPU Processing Management}
As already shown in Section~\ref{sec:scalability}, sharing a GPU among concurrent clients can add a significant overhead to the end-to-end model-serving latency. When we consider GPU resource sharing in the context of computation offload, the way GPU management can affect GPU processing is slightly different across different transport mechanisms. For instance, when using GDR, only execution engines are shared among clients.  In the case of RDMA and TCP, \emph{both} execution engines and copy engines are shared. To shed light on its implications, we evaluate different GPU management approaches and find out how effective they are in limiting GPU sharing overhead. Since TCP-based and RDMA-based transports use GPU resources similarly, we consider GDR and RDMA only.

\subsection{Concurrent Execution}
\begin{figure}
\vspace{-2ex}
\centering
    \begin{subfigure}{0.4\textwidth}
       	\centering
	   \includegraphics[width=\textwidth]{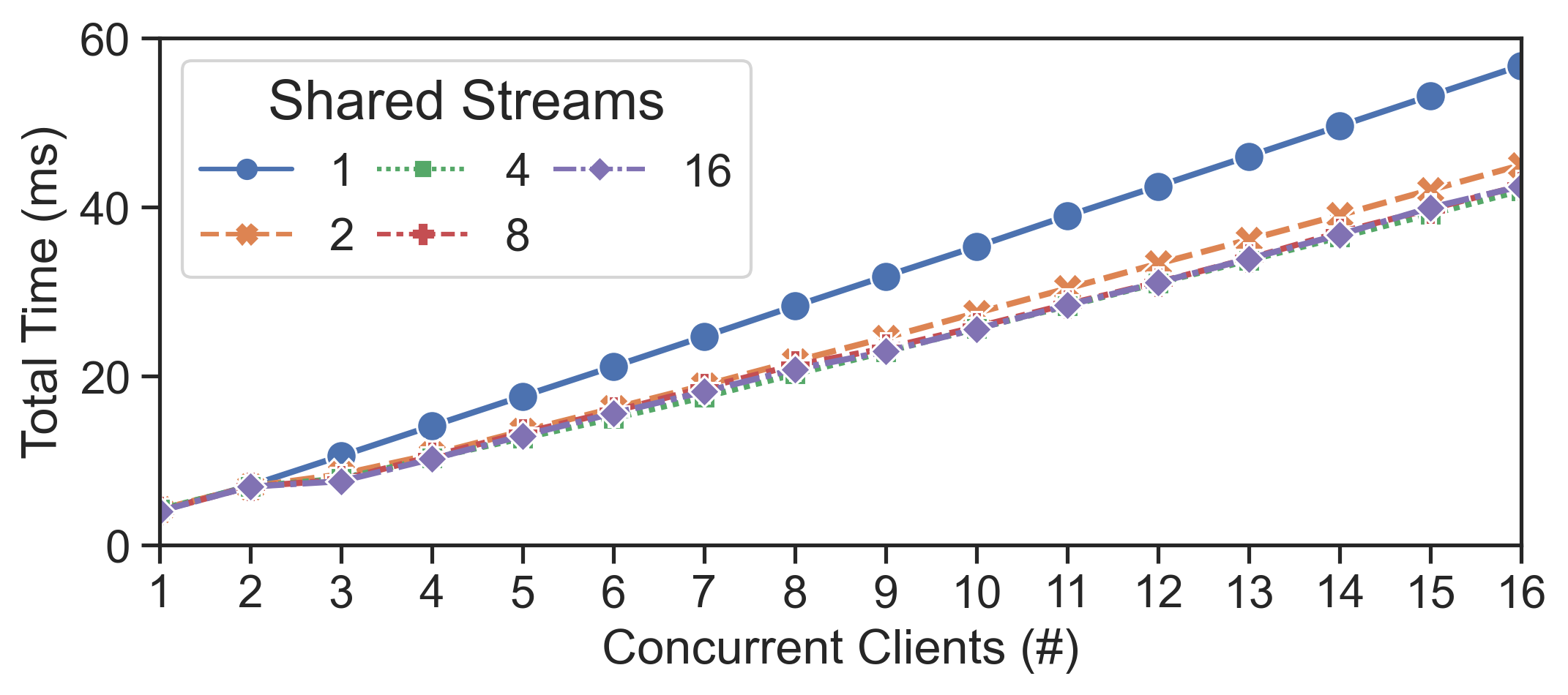}
    \vspace{-4ex}
    \caption{}
	   \label{fig:streams_resnet50_scale}
    \end{subfigure}
\\
   \begin{subfigure}{0.2\textwidth}
       	\centering
	   \includegraphics[width=\textwidth]{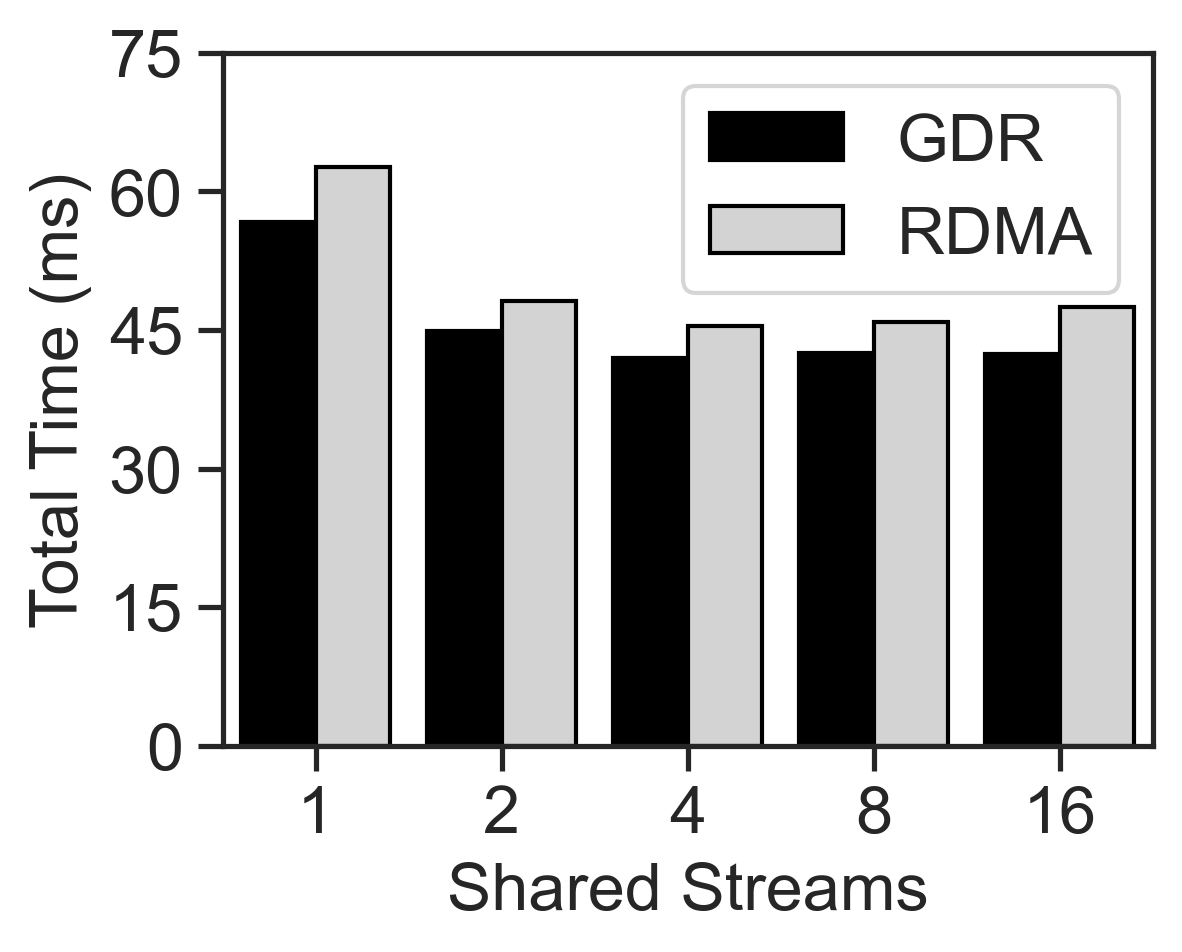}
    \vspace{-4ex}
	   \caption{}
	   \label{fig:streams_resnet50_maxclients}
    \end{subfigure}
\hspace{0.1ex}
    \begin{subfigure}{0.2\textwidth}
        \centering
        \includegraphics[width=\textwidth]{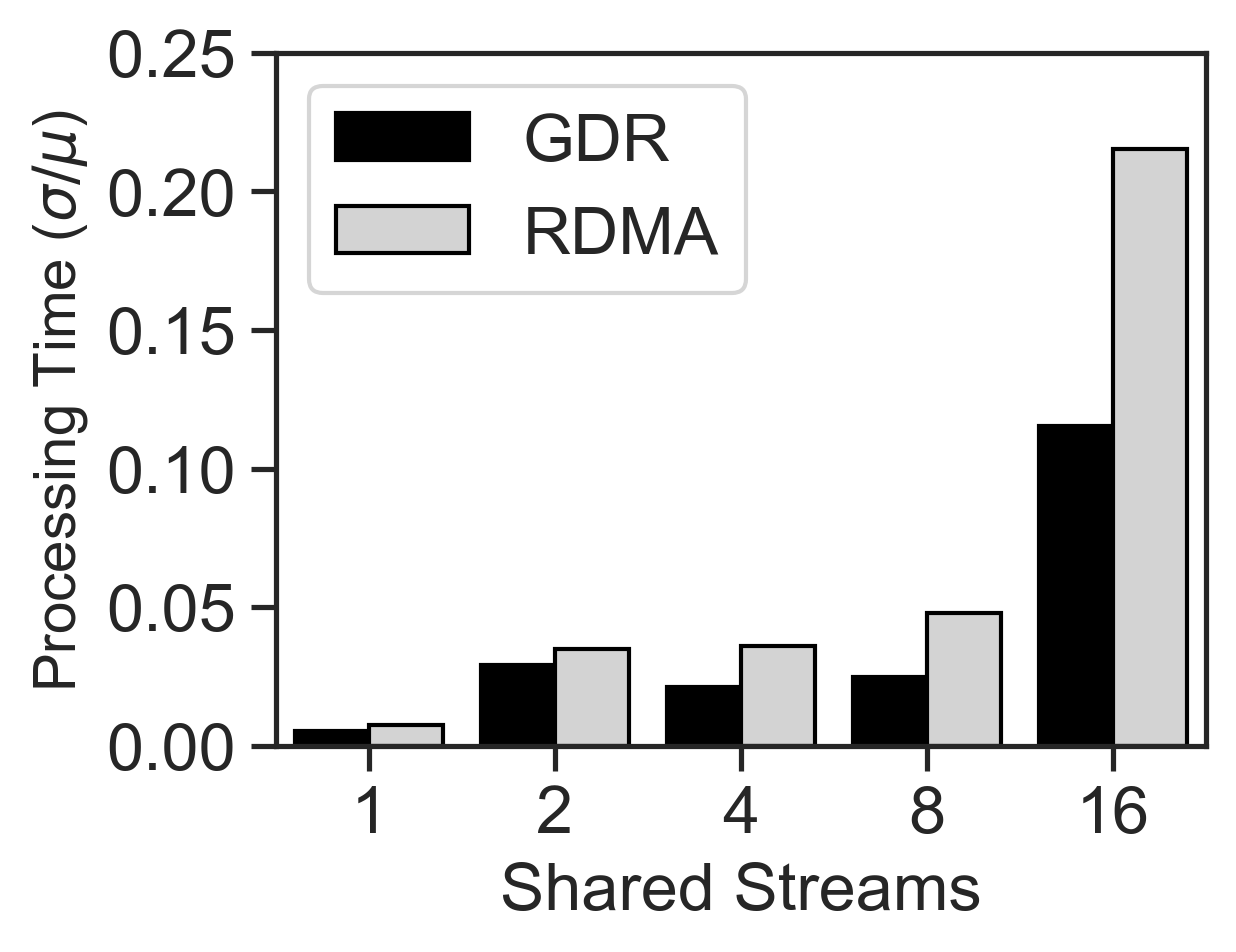}
    \vspace{-4ex}
        \caption{}
    \label{fig:streams_resnet50_processing_cov}
    \end{subfigure}
\vspace{-2ex}
\caption{Effect of limiting concurrent execution while serving \texttt{ResNet50}. (a) Scalabality using GDR, (b) Total latency when serving 16 clients, (c) CoV in processing time.
}%
\vspace{-5ex}
\label{fig:streams_resnet50}
\end{figure}

Sharing GPU resources among multiple clients increases device efficiency, but at the cost of increased variability in processing time~\cite{clockwork}. A common approach to reducing this variability is to contain the level of execution concurrency. For GPUs, this can be achieved through limiting the number of execution streams. With a fixed number of streams, client requests are placed in a job queue till a stream becomes available. This presents a trade-off between efficiency and variability regarding concurrent execution. Here we try to quantify the trade-off with different spans of GPU concurrency, i.e., concurrency in execution engines only (with GDR) and concurrency in execution and copy engines (with RDMA). 

Figure~\ref{fig:streams_resnet50} shows the effect of limiting the level of concurrency by scheduling clients with a varying number of streams on \texttt{ResNet50}. Figure~\ref{fig:streams_resnet50}(a) shows how model-serving scales when using GDR with different levels of concurrency. When there is only one stream to be shared by all clients (no-concurrent execution), end-to-end latency is 33\% higher compared to the one-stream-per-client case (i.e., 16 streams for max concurrency).  This is because limiting the number of execution streams increases requests' queuing delay. 
Figure \ref{fig:streams_resnet50}(b) zooms in on the total time for 16 clients under different levels and spans of concurrency. As the number of streams increases, both GDR and RDMA yield latency reduction, but at a monotonically decreasing rate. This is because the level of multiplexing is limited by the model and device sizes. Since GDR skips the copy engine, it performs better than RDMA. 

In Figure \ref{fig:streams_resnet50}(c), we examine the effect of concurrency by quantifying variability in processing time with the coefficient of variation (CoV) computed as $\sigma/\mu$. In both RDMA and GDR, as expected, the processing time (which excludes copy process) is less variable when concurrency is limited. However, surprisingly, the variability is different between RDMA and GDR. For instance, for 16 clients, the CoV is 0.11 and 0.21 for GDR and RDMA, respectively. This contradicts with the assumed independence between execution and copy engines in GPUs. This is mostly an artifact of the fact that GPUs are managed using a single central unit (i.e., GigaThread Engine).

\begin{figure}
\centering
    \begin{subfigure}{0.24\textwidth}
       	\centering
	   \includegraphics[width=\textwidth]{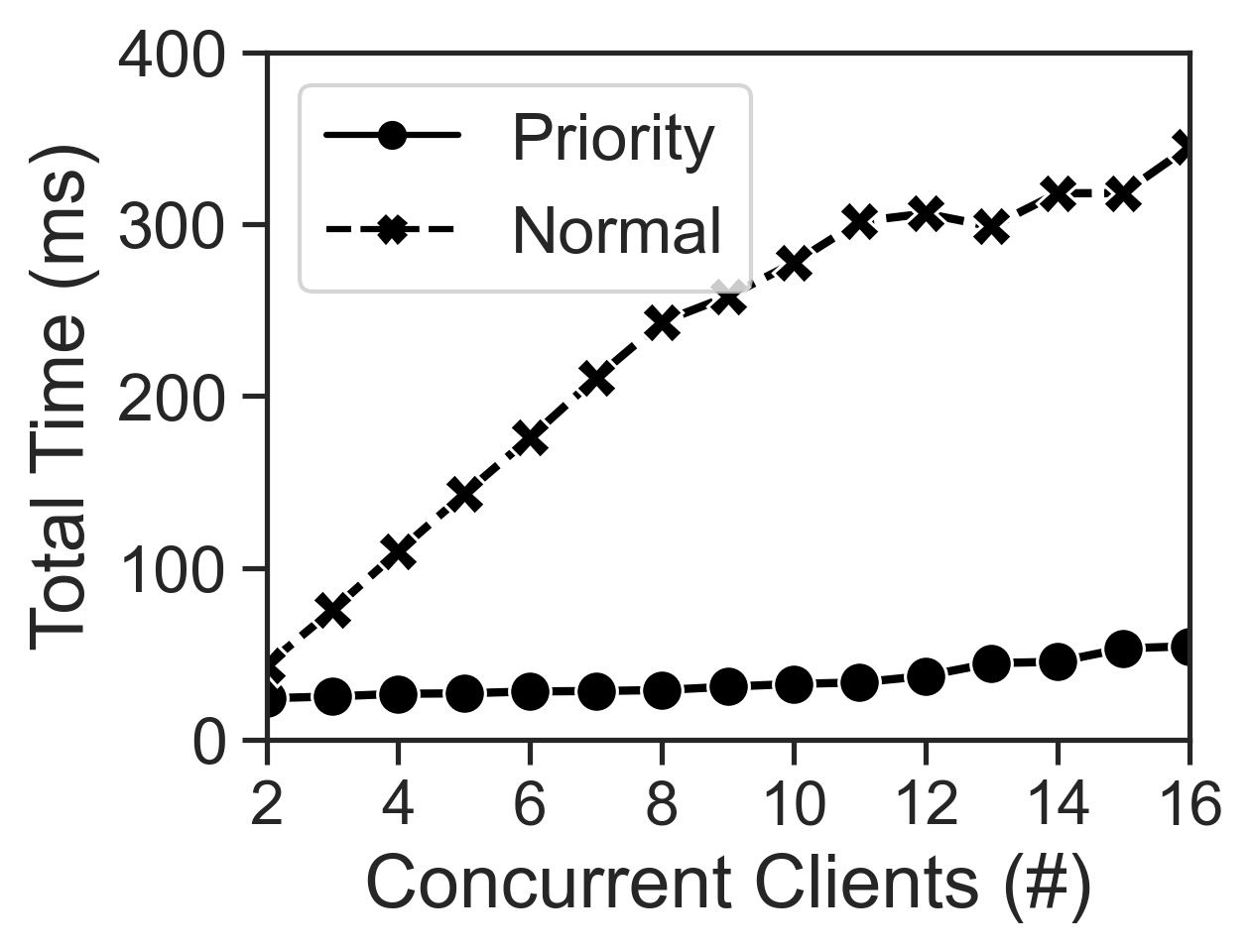}
        \vspace{-2ex}
	   \caption{GDR}
	   \label{fig:pri_1_yolov4_processed_gpudirect}
    \end{subfigure}
\hfill
    \begin{subfigure}{0.24\textwidth}
        \centering
        \includegraphics[width=\textwidth]{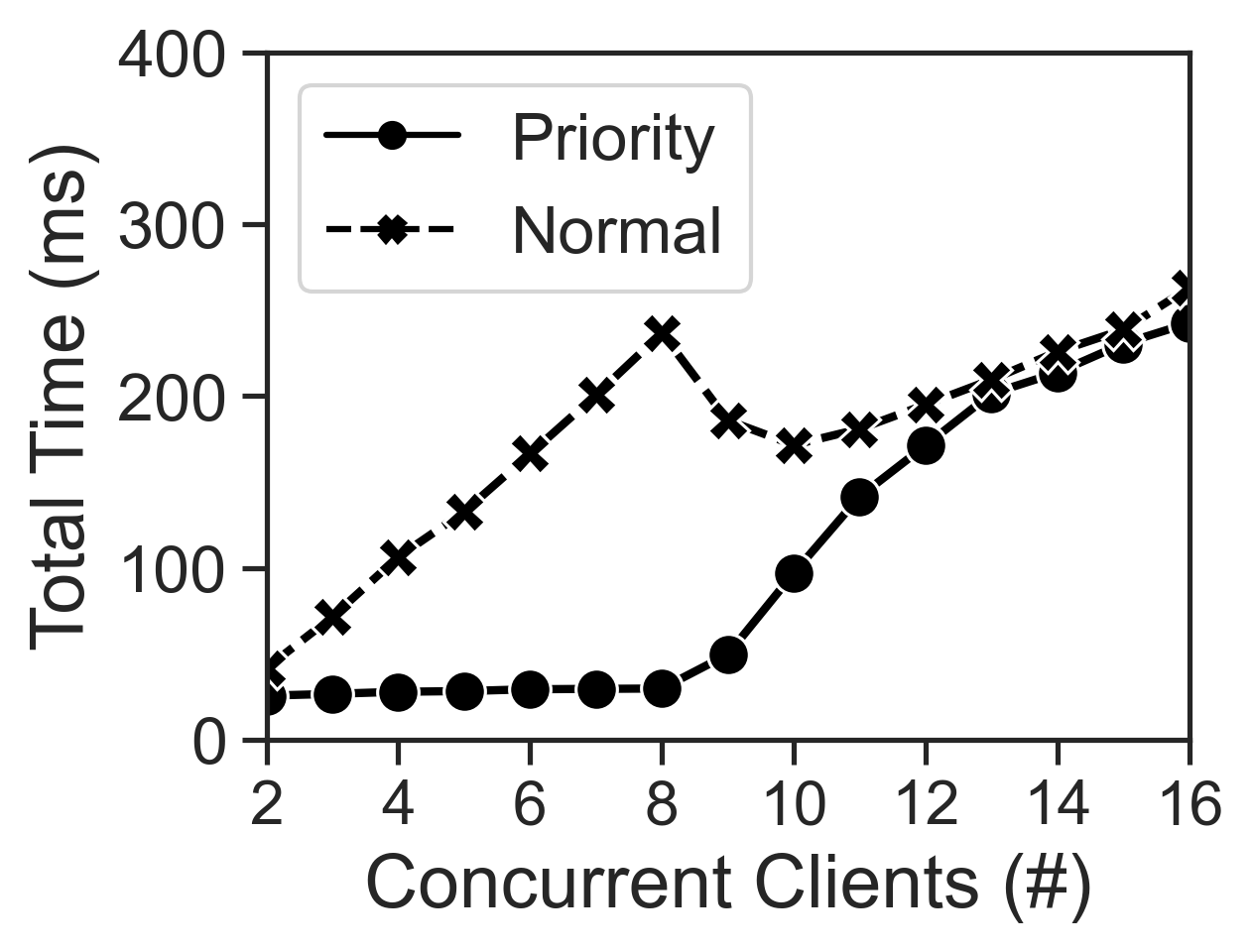}
        \vspace{-2ex}
        \caption{RDMA}
\label{fig:pri_1_yolov4_processed_rdma}
    \end{subfigure}            
\vspace{-4ex}
\caption{Single priority client with a varying number of regular clients, serving preprocessed images with \texttt{YoloV4}.}%
\vspace{-2.8ex}
\label{fig:pri_1_yolov4_preprocessed}
\end{figure}

\begin{figure}[t]
\centering
\begin{subfigure}{0.3\textwidth}
        \raggedright
       \includegraphics[width=\textwidth]{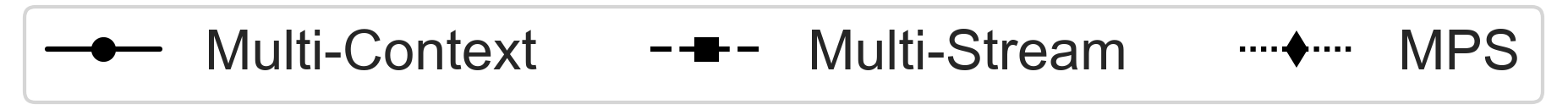}
    \end{subfigure}
    \\    
    \begin{subfigure}{0.24\textwidth}
       	\centering
	   \includegraphics[width=\textwidth]{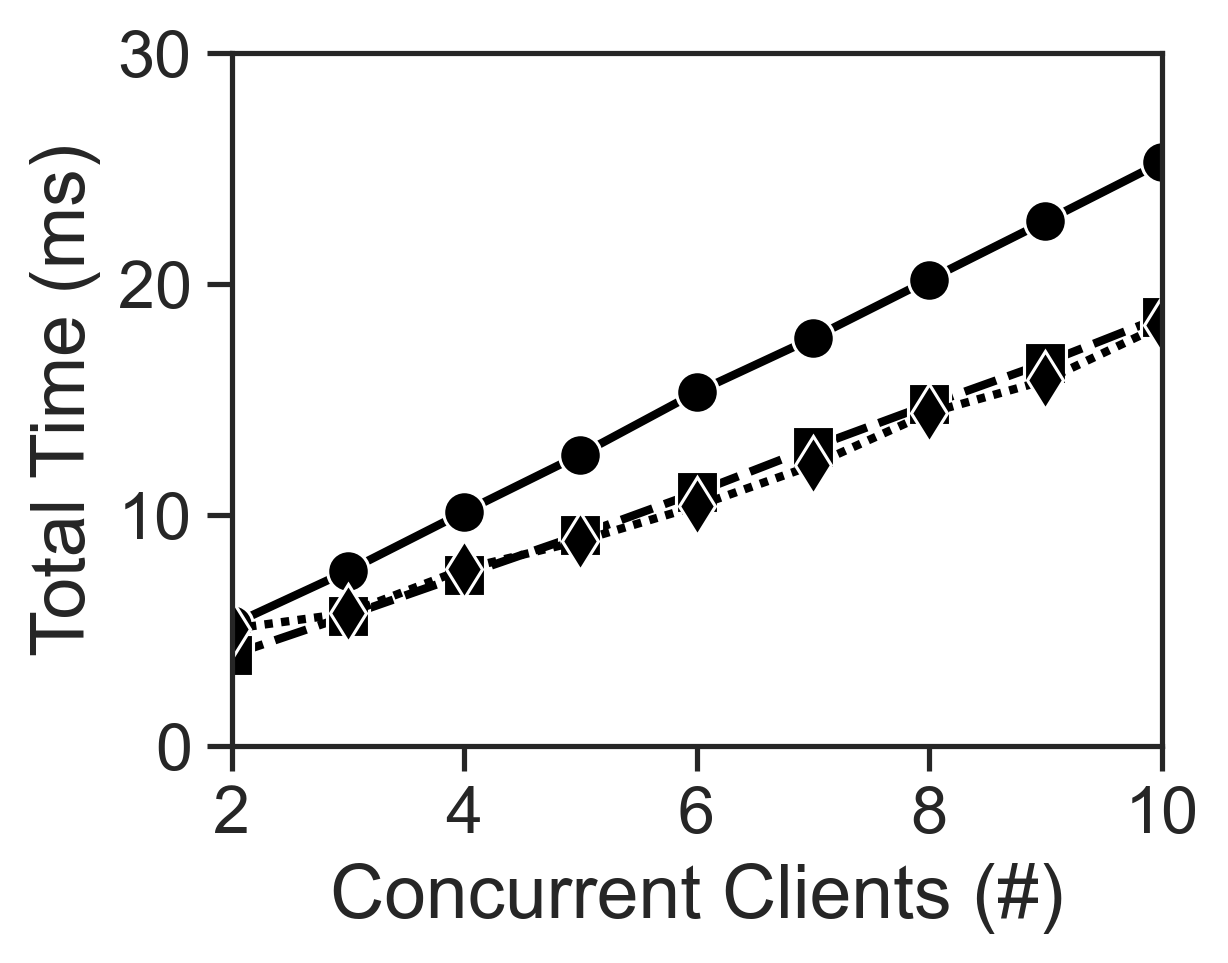}
        \vspace{-2ex}
	   \caption{GDR}
\label{fig:scheduling_raw_gpudirect_efficientnet_b0}
    \end{subfigure}
\hfill
    \begin{subfigure}{0.24\textwidth}
        \centering
        \includegraphics[width=\textwidth]{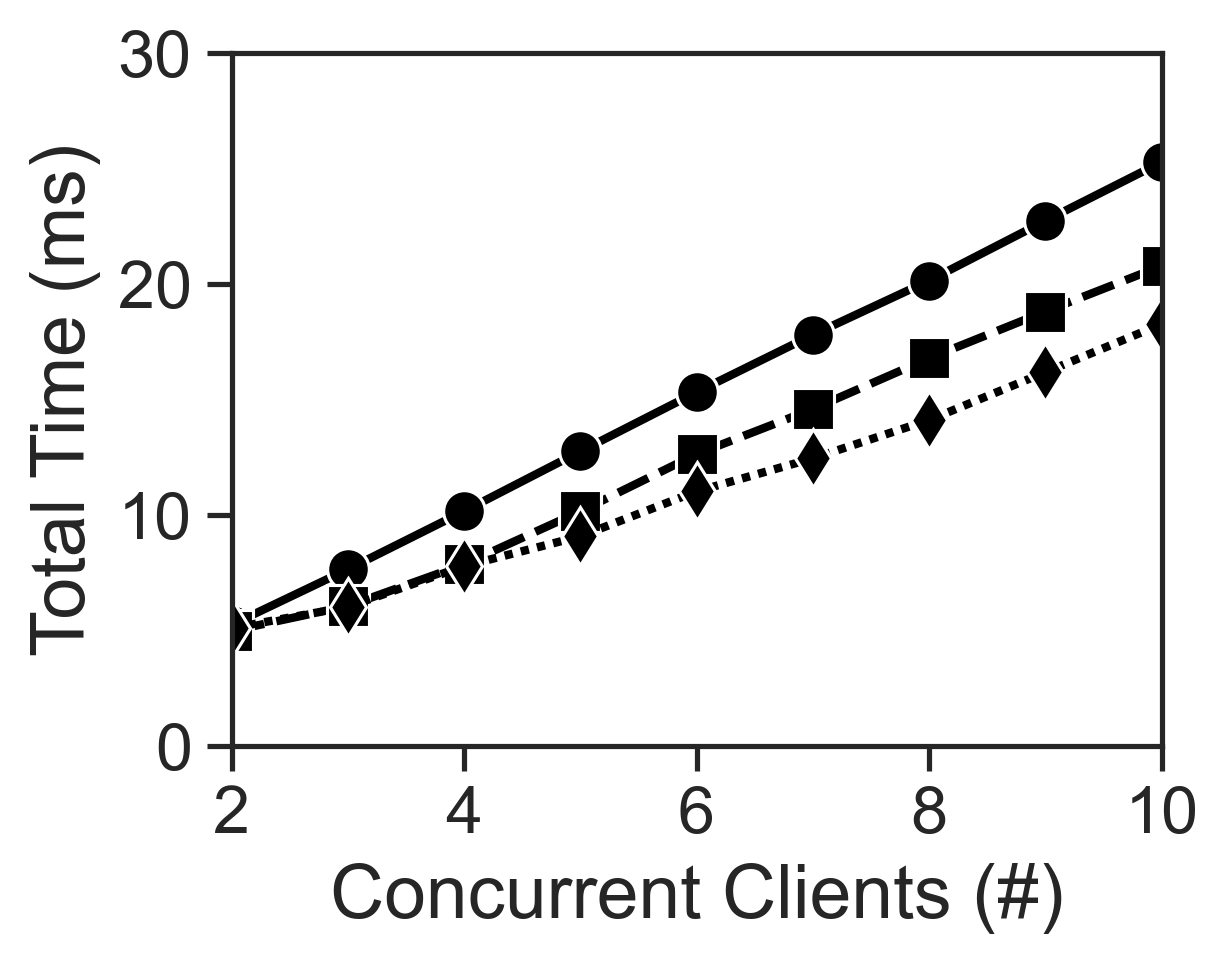}
        \vspace{-2ex}
        \caption{RDMA}
        \label{fig:scheduling_raw_rdma_efficientnet_b0}
    \end{subfigure}  
\vspace{-4ex}
\caption{Serving raw images using \texttt{EfficientNetB0} with different GPU sharing methods.}%
\vspace{-4ex}
\label{fig:scheduling_raw_efficientnet_b0}
\end{figure}


\subsection{Priority Clients}
Next, we evaluate the effect of varying the priority of different clients under two different concurrency domains (i.e., GDR and RDMA). 
In this experiment, we run one high-priority client along with other normal-priority clients, and vary the total number of clients from 2 to 16.
Figure~\ref{fig:pri_1_yolov4_preprocessed} compares model-serving latency experienced by the priority and normal clients across GDR and RDMA.  We use \texttt{YoloV4} with preprocessed images. For both RDMA and GDR, the latency for the priority client remains roughly the same with the number of clients until it reaches eight clients. In this case, the performance of the priority client greatly differs between RDMA and GDR.  
For instance, using GDR, the latency experienced by the priority client is \SI{54}{\ms} which is much lower than normal clients. However, in the case of RDMA, with more clients, the priority client's latency grows comparable to that of normal clients. This noticeable difference stems from the fact that, when using GDR, the priority is applicable only to execution that can be prioritized at a fine granularity of kernel block level~\cite{tx2scheduling2017}. In contrast, when using RDMA, the copy engine is interleaved at a coarse granularity of a request, limiting the ability of the priority client to occupy the copy engine. Note that, for models with smaller I/O, data copies will be interleaved at a relatively finer granularity, decreasing this effect on high-priority clients.

\subsection{Comparison of GPU Sharing Methods}
Lastly, we compare several common GPU sharing methods under different concurrency domains. 
We adjust the levels of concurrency by (i) changing the number of streams (\texttt{multi-stream}), (ii) increasing the number of deployed application instances with multiple contexts (\texttt{multi-context}), or (iii) time-sharing multiple application instances via \texttt{MPS}, where no streams or applications are shared between users. 
Figures~\ref{fig:scheduling_raw_efficientnet_b0}(a) and (b) compare model-serving latency among these three schemes when serving \texttt{EfficientNetB0} over GDR and RDMA. As expected, \texttt{MPS} always performs better than \texttt{multi-context} \cite{mps}. GDR and RDMA yield similar latency with \texttt{multi-context} and \texttt{MPS} across a varying number of clients, showing no clear benefit for GDR compared to RDMA. 
On the other hand, \texttt{multi-stream} exhibits different behavior between the two. As in Figure~\ref{fig:scheduling_raw_efficientnet_b0}(a), when using GDR, the performance is almost identical between \texttt{multi-stream} and \texttt{MPS}, but using RDMA, \texttt{MPS} shows better performance. We hypothesize that, across processes, GPU copy engines are interleaved in a different way, which hides the copy overhead. We note that \texttt{multi-stream} uses multi-threading which shares the CUDA libraries on the GPU,  while \texttt{multi-context} and \texttt{MPS} use multi-processing which restricts memory sharing, and hence limits the number of clients.

\noindent\emph{\textbf{Key takeaways:}
Data exchange between the host and GPU memory imposes an interfering effect on processing. 
Stream priorities are more effective in sharing execution engines than the copy engine as the scheduling decision for execution engines is made on a fine granularity. 
The copy engine is shared differently between multiple streams and contexts.
}

\section{Limitations of RDMA and GDR}\label{sec:discussion}
We show that RDMA and GDR are promising alternatives to TCP-based transport for latency-sensitive compute offloads. However, we acknowledge their drawbacks as follows.\\
\noindent\textbf{Memory overhead:} With RDMA and GDR, it is common that memory buffers are reserved and pinned per-client. This implies that the total number of sessions that can be supported will be limited, especially for GDR, as GPU memory is often more limited that host memory.\\
\noindent\textbf{Homogeneity:} RDMA transfers raw bytes between a client and a server. This requires data are stored homogeneously on both sides' memory, which might limit the interoperability. Proxied communication could alleviate this problem. 
\\
\noindent\textbf{GPU pinning:} GDR operates by allocating GPU memory buffers for each client. This ties a client to a specific GPU, or forces it to pay the data copying overhead between GPUs.  \\
\noindent\textbf{GPU inadequacy:} With dedicated ASIC-based accelerators (e.g., image decoder), GPUs may not be an optimal choice for certain preprocessing tasks, where transferring data to the host memory via RDMA may be a better option than GDR. GPUDirect may still be used to move preprocessed data from an accelerator to a GPU directly, avoiding multiple data copies.

\section{Related Work}\label{sec:related}

The performance of hardware-accelerated transports has been studied previously \cite{rdma_evaluation_13, rdma_evaluation_11, rdma_evaluation_09, gpu_interconnect2020, broadcast2014, multicast2020}. In contrast, we performed detailed performance evaluation of model-serving systems across a wide range of realistic scenarios. The most related works are \cite{lynx, gpu-ether, flex-driver}.
Lynx~\cite{lynx} offloads the network stack from CPUs to SmartNIC cores. FlexDriver~\cite{flex-driver} leverages FPGA to build an on-accelerator hardware data-plane driver. GPU-Ether~\cite{gpu-ether} implements native network I/O on GPUs themselves. While these works demonstrate the performance benefits of their point solutions, these efforts do not provide in-depth insights into the role of hardware-accelerated transport such as GDR actually plays. Our study bridges the gap and sheds light on this topic with important findings beyond simply showing which communication mechanism is the best. 

\section{Conclusion}
In this paper, we presented a reference model-serving application framework with support for multiple communication mechanisms (TCP, RDMA, GDR) and the capability to profile a model-serving pipeline on a fine time granularity. 
Our evaluation results indicate that hardware-accelerated communication provides the most improvement when communication takes up a significant portion of the pipeline. Adopting hardware-accelerated communication within the compute cluster can significantly reduce latency compared to TCP-based pipelines. Our study also highlights that data copies and concurrent compute sharing can affect latency, and these insights can be used for better utilization of hardware-accelerated communication in various applications.

\vspace{-2ex}
\section*{Acknowledgement}The work was started while Walid was a summer intern at Nokia Bell Labs. Walid and Prashant were partly supported by NSF grants 2105494, 1908536, 2211302, and 2211888.

\bibliographystyle{ieeetr}
\bibliography{bibliography}

\end{document}